\documentclass[aps,pre,twocolumn,amsmath,amssymb]{revtex4-2}

\usepackage{mathrsfs}
\usepackage{graphicx}
\usepackage{adjustbox}
\usepackage{longtable}
\usepackage{booktabs}
\usepackage{bm}
\usepackage{threeparttable}
\usepackage{amsmath,amssymb}
\usepackage{color}
\usepackage{epstopdf}
\usepackage{microtype}
\usepackage{tikz}
\usepackage{mathtools}
\usepackage{comment}
\usepackage[colorlinks=true,linkcolor=blue,anchorcolor=blue, citecolor=blue,urlcolor=blue]{hyperref}

\begin{document}

\preprint{APS/123-QED}

\title{An informational interpretation of the reverse nonequilibrium molecular dynamics}

\author{Yuxing Jiao}
\email{jiaoyuxing22@mails.ucas.ac.cn}
\affiliation{Beijing National Laboratory for Condensed Matter Physics and Laboratory of Soft Matter Physics, Institute of Physics, Chinese Academy of Sciences, Beijing 100190, China}\affiliation{School of Physical Sciences, University of Chinese Academy of Sciences, Beijing 100049, China}

\author{Mingcheng Yang}
\email{mcyang@iphy.ac.cn}
\affiliation{Beijing National Laboratory for Condensed Matter Physics and Laboratory of Soft Matter Physics, Institute of Physics, Chinese Academy of Sciences, Beijing 100190, China}\affiliation{School of Physical Sciences, University of Chinese Academy of Sciences, Beijing 100049, China}

\begin{abstract}
    Reverse nonequilibrium molecular dynamics (RNEMD) is an efficient simulation method for determining transport coefficients. In this method, a flux is imposed by a seemingly artificial nonlocal exchange operation and the corresponding thermodynamic gradient is induced as a result, apparently reversing the usual force–flux causality. Although this method is well studied as a simulation technique, developing an information-thermodynamic interpretation of its Maxwell-demon-type operation and the associated apparent reversal of force–flux causality would provide a solid physical foundation for algorithms of this type. Here, we develop an information-reservoir interpretation for a representative RNEMD scheme for determining self-diffusivity. We show that the statistical bias of the information reservoir defines an information affinity that sustains a concentration gradient and a diffusive flux in the steady state. We further derive the linear transport coefficient corresponding to the information affinity in the linear response regime. Using the large deviation functional for the steady-state densities and currents, we also establish a fluctuation theorem for entropy production and the thermodynamic uncertainty relation linking entropy production to the fluctuations of the particle current. Computer simulations are performed to corroborate our theoretical results.
\end{abstract}

\maketitle

\section{Introduction}
Nonequilibrium molecular dynamics (NEMD) is an efficient and well-established simulation method for determining transport coefficients. As in experiments, the system is driven into a nonequilibrium steady state, and transport coefficients are determined from the constitutive relations between corresponding thermodynamic forces and fluxes~\cite{EvansMorriss2008}. Various simulation schemes can be devised to reproduce the same macroscopic transport behavior. In particular, common NEMD methods can be broadly divided into two classes according to whether the thermodynamic force or the flux is imposed. In the first class of NEMD, the system is perturbed out of equilibrium by imposing a prescribed thermodynamic force on it, and the resulting flux is measured in the steady state. Examples include the SLLOD algorithm for generating shear flow~\cite{EvansMorriss2008}, and dual control volume grand canonical molecular dynamics to produce a concentration gradient~\cite{HeffelfingerSwol1994}. In these NEMD methods, the imposed forces naturally result in the fluxes, consistent with the usual force–flux causality.

In contrast, in the second class, a flux is first imposed through some artificially designed nonlocal microscopic dynamics, and the corresponding thermodynamic force develops in response. The usual force–flux causality therefore appears to be reversed, and we refer to these approaches as reverse methods. Examples include the Norton-ensemble methods~\cite{EvansMorriss2008} and the reverse nonequilibrium molecular dynamics (RNEMD)~\cite{Muller1997,MullerPlathe1999,MullerPlatheBordat2004}. Although the validity of these methods for determining transport coefficients has been established, a question remains: How can this apparent reversal of causality induced by the artificial operation be understood physically? 

In this paper, to address this question in a concrete setting, we introduce a simple nonlocal label-exchange scheme, constructed in the spirit of RNEMD, for determining self-diffusivity. We consider a single-component system confined in a channel, in which each particle is initially labeled A or B with equal probability. In this scheme, we define two slab regions near the left and right boundaries. Then, at a constant rate, an A-labeled particle in the left slab and a B-labeled particle in the right slab are randomly selected, and their labels are exchanged. As a result of this nonlocal directional exchange, a stationary concentration difference between the two slabs and hence a concentration gradient across the system are established without invoking the conventional physical driving force that sustains it.

Similar selective exchange operations are also used in other RNEMD methods to generate shear flows or temperature gradients~\cite{MullerPlatheBordat2004}. These operations are reminiscent of Maxwell's demon, a being that can also generate a flux without an externally imposed thermodynamic force. Originally proposed as a challenge to the second law of thermodynamics~\cite{Maxwell}, Maxwell’s demon subsequently stimulated research into the thermodynamics of information. Within the modern thermodynamics of information, the demon’s operation can be reconciled with the second law by accounting for the thermodynamic contributions of memory and information processing~\cite{ParrondoHorowitzSagawa2015}. Specifically, frameworks such as measurement and feedback control~\cite{SagawaUeda2010,SagawaUeda2012,HorowitzEsposito2014} and information reservoirs~\cite{MandalJarzynski2012,DeffnerJarzynski2013,BaratoSeifert2014PRL,BaratoSeifert2014PRE} are proposed to incorporate the information processing into thermodynamics and provide corresponding generalized second laws. These developments suggest that applying the formalisms of information thermodynamics to reverse methods may help clarify the apparent reversal of force–flux causality and thus establish a solid physical foundation for such Maxwell-demon-type simulation techniques.

Motivated by the above observation, we interpret the generalized label-exchange RNEMD scheme within an information-reservoir framework. In a generic information-reservoir model, a thermodynamic system is coupled to a tape carrying a sequence of bits (a sequence of zeros and ones), which can act as an autonomous Maxwell's demon, extracting work from the system at the cost of increasing its own Shannon entropy~\cite{MandalJarzynski2012,DeffnerJarzynski2013,BaratoSeifert2014PRL,BaratoSeifert2014PRE}. This tape is regarded as an information reservoir because it can induce transitions between system states, as a thermal reservoir does, while affecting only the system’s entropy balance and exchanging no energy with the system. In our information-reservoir interpretation, such a tape controls the exchange process in RNEMD. We derive a reaction–diffusion-type equation to describe the exchange and diffusion process in the system and determine the steady-state entropy production. We find that the statistical bias of the incoming bits defines an information affinity that drives the label-exchange flux and sustains the concentration difference, thereby providing a physical interpretation of the apparent reversal of force–flux causality. We then analyze the linear response to the information affinity and relate the resulting transport coefficient to the self-diffusivity. Furthermore, we construct the large deviation functional for the steady-state densities and currents and derive a fluctuation theorem and the thermodynamic uncertainty relation. Finally, we test our theoretical predictions using computer simulations.

\section{The model}
\subsection{Information-reservoir interpretation of label-exchange RNEMD}
\begin{figure}[htbp!]
    \centering
    \includegraphics[keepaspectratio, width=0.7\columnwidth]{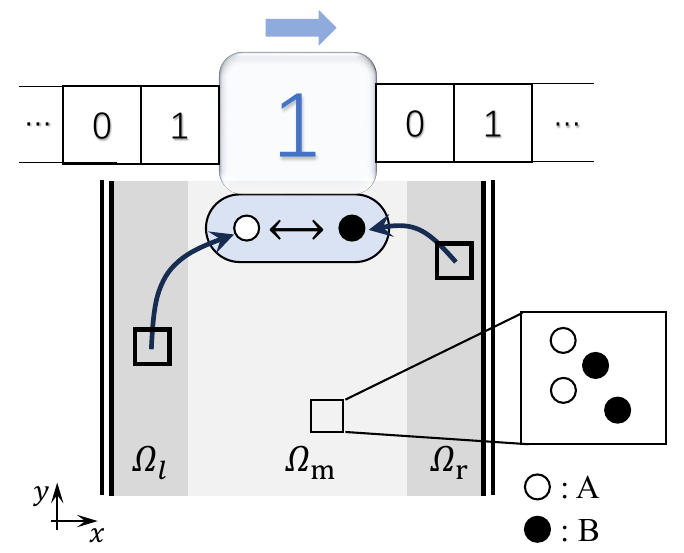}
    \caption{Schematic illustration of the system. Particles in the system are labeled A (shown as \(\circ \)) or B (shown as \(\bullet \)). The system is divided into three regions: two slabs \(\Omega_l\) and \(\Omega_r\), and a bulk region \(\Omega_m\). The system is equipped with a tape. On average, the tape moves forward by one bit every \(\langle\Delta t\rangle\), after which the bit in the window attempts to induce a label interchange between \(\Omega_l\) and \(\Omega_r\). As exemplified in the sketch, if the bit in the window is in state 1 and the selected pair of particles is \((A,B)\) (A from \(\Omega_l\) and B from \(\Omega_r\)), then the bit flips and the labels of the two particles are interchanged: \((A,B)\to(B,A)\). An analogous update applies in the opposite case.}\label{Fig::Sketch}
\end{figure}

We consider a two-dimensional, one-component fluid consisting of \(N\) particles in a square domain of side length \(L\), $\Omega=\{(x,y)\mid x,y\in[0,L]\}$, in which $N_A$ particles are labeled A and $N_B$ are labeled B (see the sketch in Fig.~\ref{Fig::Sketch}). The system is confined in the $x$-direction by walls at $x=0$ and $x=L$, and periodic boundary conditions are imposed in the \(y\) direction. We divide the system into three regions by $\Omega=\Omega_l\cup\Omega_m\cup\Omega_r$, where $\Omega_l=\{(x,y)\in\Omega\mid x\in[0,a]\}$ and $\Omega_r=\{(x,y)\in\Omega\mid x\in[L-a,L]\}$ are the left and right slabs, respectively, each of width \(a\ll L\), and $\Omega_m=\{(x,y)\in\Omega\mid x\in(a,L-a)\}$ is the bulk region. 

In the spirit of the information-reservoir model, we introduce a tape carrying a sequence of bits to control the label exchanges in RNEMD (see Fig.~\ref{Fig::Sketch}). Hereafter, we denote the labels of the particle pair selected from the two slabs $\Omega_l,\,\Omega_r$ by $(\xi_l,\xi_r)\in\{(A,B),(B,A)\}$, where $\xi_l$ and \(\xi_r\) are the labels of the particles from the left slab \(\Omega_l\) and the right slab \(\Omega_r\), respectively. The tape interacts with the system through a one-bit-wide window. After each waiting time \(\Delta t\), the tape advances by one bit position, bringing a new bit into the interaction window and mediating a label exchange. This bit interacts with the selected particle pair according to the following exchange rules. If the incoming bit is 1 and the selected pair is \((A,B)\), then the labels are interchanged (\((A,B)\to(B,A)\)) and the bit flips from 1 to 0; if the incoming bit is 0 and the selected pair is \((B,A)\), then the labels are interchanged (\((B,A)\to(A,B)\)) and the bit flips from 0 to 1. For other cases, no exchange is performed. We set the initial fractions of zeros and ones to \(\epsilon\) and \(1-\epsilon\), respectively. Consequently, if \(\epsilon\neq\frac{1}{2}\), a difference in the concentrations of A and B particles will be established between the two slabs. For the sake of theoretical simplicity, we assume the time interval \(\Delta t\) follows an exponential distribution and thus the exchange attempts form a Poisson process. We denote the corresponding Poisson rate by \(\gamma=\frac{1}{\left\langle\Delta t\right\rangle}\). If we choose \(\epsilon=1\) or \(\epsilon=0\) and take \(\Delta t\) to be a fixed time interval, this generalized RNEMD reduces to the traditional RNEMD scheme.

\subsection{The simulation model}
To realize the above generalized label-exchange RNEMD in computer simulations, we use stochastic rotation dynamics (SRD) to model the dynamics of the fluid particles. SRD is a highly efficient particle-based mesoscale simulation model for classical fluids~\cite{MPC1,Gompper_2009}. Although the interactions between the fluid particles are coarse-grained in the SRD model, it correctly captures the hydrodynamic behavior, dissipation, and thermal fluctuations of the ordinary fluids~\cite{Gompper_2009,SRD_GK1}. Therefore, it has not only been used to study the dynamics of complex fluids (as a coarse-grained solvent)~\cite{MPC_MD}, but also for studying the statistical physics of fluids, such as the transport phenomena~\cite{Luo2020} and fluctuation theorem~\cite{BelushkinLiviFoffi2011}.

The SRD fluid particles follow a discrete-time dynamics with time step \(h\). Each particle has the same mass \(m\). In each SRD step, the position \(\bm{r}_i\) and the velocity \(\bm{v}_i\) of the particle \(i\) are updated by two successive substeps, referred to as the streaming step and the collision step, respectively. In the streaming step, the particle moves ballistically according to:
\begin{equation}
    \bm{r}_i(t+h)={\bm{r}}_i(t) + \bm{v}_i(t)h.
\end{equation}
In the collision step, the particles are sorted into square lattice cells of size \(l\times l\) according to their positions. The origin of this square lattice is randomly shifted in every SRD step to preserve Galilean invariance~\cite{MPC_RS}. Then the velocity of particle \(i\) is updated by rotating its velocity relative to the center-of-mass velocity of its cell:
\begin{equation}
    \bm{v}_i(t+h)=\bm{v}_{\text{cm}} + \bm{R}(\hat{\omega})\cdot(\bm{v}_i(t)-\bm{v}_{\text{cm}}),
\end{equation}
where \(\bm{v}_{\text{cm}}\) is the center-of-mass velocity of the cell containing particle \(i\), \(\bm{R}(\hat{\omega})\) is the two-dimensional rotation operator, and \(\hat{\omega}\) is the rotation angle, chosen from \(\{\omega,-\omega\},\omega\in(0,\pi)\) with equal probability. This dynamics exactly conserves particle number, cell-level momentum, and energy. SRD can quickly relax to an equilibrium state (microcanonical distribution) and satisfies an ideal gas equation of state in the absence of external perturbations.

In RNEMD-SRD simulations, we set \(l=1,\,m=1,\,h=0.1,\,\text{mean thermal energy } k_{\text{B}}T=1\), and \(L=20,\rho=\frac{N}{L^2}=20,N_A=N_B=\frac{1}{2}N\).

\subsection{Hydrodynamic description}
In this section, we derive a reaction-diffusion-type hydrodynamic equation to describe this RNEMD scheme. We denote the densities of A and B particles by \(\rho_A\) and \(\rho_B\), respectively. Because the label-exchange operation conserves the total number of particles, \(N=N_A+N_B\), the total density \(\rho=\rho_A+\rho_B\) is constant. This allows us to use a single equation for \(\rho_A\) to describe the system. Owing to translational symmetry in the \(y\) direction, we take the hydrodynamic fields to be independent of \(y\). We also temporarily omit the time variable \(t\). For example, we write \(\rho_A(\bm{r},t)\equiv\rho_A(x)\). We define \(j_A^\psi(\bm{r},\bm{r}')\) as the local label-exchange current density of A particles flowing from the right slab \(\bm{r}'\in\Omega_r\) to the left slab \(\bm{r}\in\Omega_l\). According to the symmetry, we simply denote \(\text{j}_A^\psi(x,x')=\int_0^L\text{d}y\int_0^L\text{d}y'j_A^\psi(\bm{r},\bm{r}')=L^2j_A^\psi(\bm{r},\bm{r}')\)~\footnote{Hereafter, the upright symbols such as \(\text{j}_A^\psi(x,x')\) here and \(\text{k}_{\pm}(x,x')\) defined below denote exchange-current densities per pair of line elements $\text{d}x\,\text{d}x'$. The italic symbols such as \({j}_A^\psi(\bm{r},\bm{r}')\) and \({k}_{\pm}(\bm{r},\bm{r}')\) in the later section denote exchange-current densities per pair of area elements $\text{d}\bm{r}\,\text{d}\bm{r}'$.}. The density \(\rho_A\) in the left and right slabs evolves according to the following two reaction-diffusion-type equations:
\begin{equation}\label{RDeq}
    \begin{aligned}
        &\partial_t\rho_A = D\partial_x^2 \rho_A  + \frac{1}{L}\int_{L-a}^{L}\text{d}x'\text{j}_A^\psi(x,x') ,\qquad x\in[0,a],\\
        &\partial_t\rho_A = D\partial_x^2 \rho_A  - \frac{1}{L}\int_{0}^{a}\text{d}x'\text{j}_A^\psi(x',x) , \qquad x\in[L-a,L],
    \end{aligned}
\end{equation}
where \(D\) is the self-diffusivity. In the bulk region, \(\rho_A\) is simply described by the diffusion equation
\begin{equation}\label{BulkDeq}
    \partial_t \rho_A = D\partial_x^2 \rho_A,\qquad x\in(a,L-a).
\end{equation}
We remark that this hydrodynamic description is valid under the following conditions: 1) the width of the slabs is much larger than the mean free path of the fluid particles, and 2) the diffusive current lies within the linear-response regime, which requires that the label exchange not be too frequent.

We denote the net label-exchange current of A particles flowing from \([L-a,L]\) into \([0,a]\) by
\begin{equation}\label{jApsi0}
    \text{j}_A^\psi(x,x') = \text{k}_+(x,x') - \text{k}_-(x,x'),
\end{equation}
where \(x\in[0,a], x'\in[L-a,L]\), and \(\text{k}_+(x,x')\) and \(\text{k}_-(x,x')\) are the currents flowing into and out of the left slab \(\Omega_l\), respectively. Specifically, \(\text{k}_+(x,x')\) corresponds to the case in which the incoming bit is 0 and the selected pair \((\xi,\xi')\) is \((B,A)\), whereas \(\text{k}_-(x,x')\) corresponds to the label-change case in which the incoming bit is 1 and the selected pair is \((A,B)\). We thus obtain the following expressions:
\begin{widetext}
    \begin{equation}\label{jApsiDef}
        \begin{aligned}
            \text{k}_+(x,x')=&\gamma\epsilon P\left(x,\xi=B;x',\xi'=A\Bigg| x\in[0,a],x'\in[L-a,L],(\xi,\xi')\in\{(A,B),(B,A)\}\right),\\
            \text{k}_-(x,x')=&\gamma(1-\epsilon)P\left(x,\xi=A;x',\xi'=B\Bigg| x\in[0,a],x'\in[L-a,L],(\xi,\xi')\in\{(A,B),(B,A)\}\right),
        \end{aligned}
    \end{equation}
where the conditional probability density reads
\begin{equation}\label{pBAcond}
    \begin{aligned}
        &P\left(x,\xi=B;x',\xi'=A\Bigg| x\in[0,a],x'\in[L-a,L],(\xi,\xi')\in\{(A,B),(B,A)\}\right)\\
        &=\frac{p(x,B;x',A)}{\int_0^a\text{d}x\int_{L-a}^{L}\text{d}x'\left( p(x,B;x',A) + p(x,A;x',B) \right)}\triangleq\frac{p(x,B;x',A)}{Z},
    \end{aligned}
\end{equation}
and \(p(x,B;x',A)\) is the probability density for selecting a B particle at \(x\) and an A particle at \(x'\). If the distance \(|x-x'|\) is much larger than the interaction range between the particles, the two-particle distribution can be expressed in terms of the single-particle distributions \(p_A(x)=\int_0^L p_A(x,y)\text{d}y=\frac{1}{N_A}\int_0^L\rho_A(x)\text{d}y=\frac{L}{N_A}\rho_A(x)\) and \(p_B(x)=\frac{L}{N_B}\rho_B(x)\) as follows:
\begin{equation}\label{pBA}
    p(x,B;x',A)=p_B(x)p_A(x')=\frac{L^2}{N_AN_B}\rho_B(x)\rho_A(x').
\end{equation}
For the current \(\text{j}_A^\psi(x,x')\), with \(x\in[0,a]\) and \(x'\in[L-a,L]\), the separation \(|x-x'|\sim L\) is sufficiently large. We can therefore apply Eq.~\eqref{pBA} to Eq.~\eqref{jApsiDef}, which gives
\begin{equation}\label{jApsi}
    \text{j}_A^\psi(x,x') = \frac{\gamma\epsilon}{Z} \frac{L^2}{N_AN_B}\rho_B(x)\rho_A(x') - \frac{\gamma(1-\epsilon)}{Z}\frac{L^2}{N_AN_B}\rho_A(x)\rho_B(x'),\qquad Z=\frac{N_B^lN_A^r+N_A^lN_B^r}{N_AN_B},
\end{equation}
where \(N_\xi^l\) and \(N_\xi^r\) denote the numbers of \(\xi\) particles, with \(\xi=A,B\), in the left and right slabs, respectively:
\begin{equation}
    N_\xi^l=L\int_0^a\text{d}x\,\rho_\xi(x),
    \qquad
    N_\xi^r=L\int_{L-a}^{L}\text{d}x\,\rho_\xi(x).
\end{equation}
Substituting the explicit expression for \(\text{j}_A^\psi\) in Eq.~\eqref{jApsi} into Eq.~\eqref{RDeq}, and combining the result with Eq.~\eqref{BulkDeq} and the zero-flux boundary conditions at \(x=0\) and \(x=L\), we obtain the governing equation and boundary conditions for the system:
\begin{align}
    &\partial_t \rho_A = D\partial_x^2 \rho_A  - \gamma(1-\epsilon)\frac{N_B^r}{ZN_AN_B}\rho_A + \gamma\epsilon\frac{N_A^r }{ZN_AN_B} \rho_B,\qquad x\in[0,a],\label{GvrnEqL}\\
    &\partial_t \rho_A = D\partial_x^2 \rho_A,\qquad x\in(a,L-a),\label{GvrnEqM}\\
    &\partial_t \rho_A = D\partial_x^2 \rho_A  + \gamma(1-\epsilon)\frac{N^l_A}{ZN_AN_B} \rho_B - \gamma\epsilon\frac{N^l_B}{ZN_AN_B} \rho_A, \qquad x\in[L-a,L],\label{GvrnEqR}\\
    &\partial_x \rho_A|_{x=0,L}=0,\label{GvrnEqB}
\end{align}
where \(\rho_B=\rho-\rho_A\).
\end{widetext}

\section{Macroscopic steady-state entropy production rate}
Within the hydrodynamic description, we derive the macroscopic steady-state entropy production rate of the system. In the steady state, the label-exchange flux is balanced by the diffusion flux, and these two types of flux contribute separately to the macroscopic entropy production rate. We thus decompose the total entropy production rate as
\begin{equation}\label{EPRtot}
    \dot{\Sigma}_{\text{ss}}^{\text{tot}}=\dot{\Sigma}^{\text{diff}}_{\text{ss}} + \dot{\Sigma}^{\psi}_{\text{ss}}
\end{equation}
where \(\dot{\Sigma}^{\text{diff}}_{\text{ss}}\) and \(\dot{\Sigma}^{\psi}_{\text{ss}}\) denote the diffusion and label-exchange contributions, respectively, and we use ``\({\text{ss}}\)'' to label the steady-state quantities hereafter.

The diffusion entropy production rate is given by
\begin{equation}\label{EPRdiffori}
    \begin{aligned}
        \dot{\Sigma}^{\text{diff}}_{\text{ss}}&=\left( \int_{\Omega_l} + \int_{\Omega_m} + \int_{\Omega_r} \right)\text{d}\bm{r}(-{j^{\text{ss}}_A}_x\partial_x\mu_A-{j_B^{\text{ss}}}_x\partial_x\mu_B)\\
        &\triangleq \dot{\Sigma}^{\text{diff}}_{\text{ss},\Omega_l} + \dot{\Sigma}^{\text{diff}}_{\text{ss},\Omega_m} + \dot{\Sigma}^{\text{diff}}_{\text{ss},\Omega_r},
    \end{aligned}
\end{equation}
where \({j_{\xi x}^{\text{ss}}}=-D\partial_x\rho_\xi^{\text{ss}}\), with \(\xi\in\{A,B\}\), is the diffusion flux and \(\mu_\xi\) is the chemical potential of \(\xi\) particles. Here, we consider an ideal-gas or ideal-solution system, so the chemical potential is \(\mu_\xi=\ln\frac{\rho_\xi^{\text{ss}}}{\rho}\), where we have set \(k_BT=1\) because the system is isothermal. Substituting this chemical potential into Eq.~\eqref{EPRdiffori} and using the relation \(\rho_A^{\text{ss}}+\rho_B^{\text{ss}}=\rho=\text{const.}\), we obtain
\begin{equation}\label{EPRdiff}
    \dot{\Sigma}^{\text{diff}}_{\text{ss}}=-L\int_0^L\text{d}xj_{Ax}^{\text{ss}}\partial_x\ln\frac{\rho_A^{\text{ss}}(x)}{\rho_B^{\text{ss}}(x)}.
\end{equation}
According to the steady-state equation for the bulk region, Eq.~\eqref{GvrnEqM}, the diffusion flux in the bulk is constant. We therefore write
\({j_A^{\text{ss}}}_x(x)={j_A^{\text{ss,bulk}}}_x={j_A^{\text{ss}}}_x(a)={j_{Ax}^{\text{ss}}}(L-a)\) for \(x\in(a,L-a)\). The diffusion entropy production rate in the bulk region is then
\begin{equation}
    \dot{\Sigma}^{\text{diff}}_{\text{ss},\Omega_m}=L{j_{Ax}^{\text{ss,bulk}}}\ln\frac{\rho_A^{\text{ss}}(a)\rho_B^{\text{ss}}(L-a)}{\rho_A^{\text{ss}}(L-a)\rho_B^{\text{ss}}(a)}.
\end{equation}

To quantify the irreversibility of the label-exchange process, we define its macroscopic entropy production rate using the standard Schnakenberg form~\cite{Schnakenberg1976}:
\begin{equation}\label{EPRpsiori}
    \dot{\Sigma}^{\psi}_{\text{ss}}=\int_0^a\text{d}x\int_{L-a}^L\text{d}x'\,\text{j}_{A,\text{ss}}^\psi(x,x')\ln\frac{\text{k}_+^{\text{ss}}(x,x')}{\text{k}_-^{\text{ss}}(x,x')}.
\end{equation}
Using Eqs.~\eqref{jApsiDef}--\eqref{pBA}, the steady-state governing equations~\eqref{GvrnEqL}--\eqref{GvrnEqB}, and the diffusion entropy production rate in Eq.~\eqref{EPRdiff}, we simplify Eq.~\eqref{EPRpsiori} to
\begin{equation}\label{EPRpsi}
    \dot{\Sigma}^\psi_{\text{ss}}=J_{A,\text{ss}}^\psi F^\psi-\dot{\Sigma}^{\text{diff}}_{\text{ss}},
\end{equation}
in which we have introduced the total particle flux \(J_A^\psi\) generated by the label exchange and \(F^\psi\) as
\begin{equation}\label{DefJApsi}
    J_{A,\text{ss}}^\psi=\int_0^a\text{d}x\int_{L-a}^L\text{d}x'\text{j}_{A,\text{ss}}^\psi(x,x'),\qquad F^\psi=\ln\frac{\epsilon}{1-\epsilon}.
\end{equation}
Combining Eqs.~\eqref{EPRtot}, \eqref{EPRpsi}, \eqref{DefJApsi}, we find that the resulting macroscopic entropy production rate in the steady state takes the standard flux--affinity form:
\begin{equation}\label{SSEPR0}
    \dot{\Sigma}_{\text{ss}}^{\text{tot}}=J_{A,\text{ss}}^\psi F^\psi.
\end{equation}
Consequently, under our informational interpretation, the flux \(J_{A,\text{ss}}^\psi\) imposed in RNEMD is actually driven by \(F^\psi\), which originates from the statistical bias of the incoming bit sequence and represents the affinity of the information reservoir~\cite{BaratoSeifert2014PRE}.

To further derive the expression for the total flux \(J_{A,\text{ss}}^\psi\), we denote the probability that an output bit is in state \(I\in\{0,1\}\) by \(p_I\). We note that \(p_0\) and \(p_1\) correspond, respectively, to the probabilities of selecting the pairs \((A,B)\) and \((B,A)\):
\begin{equation}\label{p0p1}
    \begin{aligned}
        p_0&=\int_0^a\text{d}x\int_{L-a}^L\text{d}x'\frac{p(x,A;x',B)}{Z}=\frac{N_A^lN_B^r}{N_B^lN_A^r+N_A^lN_B^r},\\
        p_1&=\int_0^a\text{d}x\int_{L-a}^L\text{d}x'\frac{p(x,B;x',A)}{Z}=\frac{N_B^lN_A^r}{N_B^lN_A^r+N_A^lN_B^r}.
    \end{aligned}
\end{equation}
Based on Eqs.~\eqref{jApsi} and \eqref{p0p1}, the total flux under the expression for \(p_0\) reads
\begin{equation}\label{Jpsi}
    J_{A,\text{ss}}^\psi=\gamma\epsilon p_1^{\text{ss}} - \gamma(1-\epsilon)p_0^{\text{ss}}=\gamma(\epsilon - p_0^{\text{ss}}).
\end{equation}
With the help of Eqs.~\eqref{DefJApsi}, \eqref{SSEPR0} and \eqref{Jpsi}, the entropy production rate becomes:
\begin{equation}\label{SSEPR}
    \dot{\Sigma}_{\text{ss}}^{\text{tot}}=\gamma(\epsilon - p_0^{\text{ss}})\ln\frac{\epsilon}{1-\epsilon}.
\end{equation}

We emphasize that the total entropy production rate \(\dot{\Sigma}_{\text{ss}}^{\text{tot}}\) is not related to dissipation in the system because the system is not in contact with any particle or thermal reservoirs. To further investigate the role of the entropy production rate \(\dot{\Sigma}_{\text{ss}}^{\text{tot}}\), we decompose it into two nonnegative terms:
\begin{equation}\label{DcpSigma}
    \dot{\Sigma}_{\text{ss}}^{\text{tot}} = \gamma (H(p_0^{\text{ss}})-H(\epsilon)) + \gamma D_{\text{KL}}(p_0^{\text{ss}}\|\epsilon),
\end{equation}
where \(H(q)=-q\ln q-(1-q)\ln(1-q)\), with \(q\in[0,1]\), is the Shannon entropy, and \(D_{\text{KL}}(p_0^{\text{ss}}\|\epsilon)=p_0^{\text{ss}}\ln\frac{p_0^{\text{ss}}}{\epsilon}+(1-p_0^{\text{ss}})\ln\frac{1-p_0^{\text{ss}}}{1-\epsilon}\) is the Kullback--Leibler divergence. It can be shown that both terms are nonnegative, i.e., \(\gamma (H(p_0^{\text{ss}})-H(\epsilon))\geqslant 0,\,\gamma D_{\text{KL}}(p_0^{\text{ss}}\|\epsilon)\geqslant 0\)~\footnote{The inequality \(H(p_0^{\text{ss}})-H(\epsilon)\geqslant 0\) can be proved as follows. In the case of \(\epsilon\geqslant\frac{1}{2}\), the nonnegativity of \(\dot{\Sigma}_{\text{ss}}^{\text{tot}}\) implies \(\epsilon\geqslant p_0^{\text{ss}}\). The resulting total flux \(J_{A,\text{ss}}^\psi=\gamma(\epsilon - p_0^{\text{ss}})\geqslant 0\) implies more A labels and B labels are exchanged into \(\Omega_l\) and \(\Omega_r\), respectively. This means \(N_A^lN_B^r\geqslant N_B^lN_A^r\) and thus \(p_0^{\text{ss}}\geqslant\frac{1}{2}\) (cf. Eq.~\protect\eqref{p0p1}). Therefore we have \(\epsilon\geqslant p_0^{\text{ss}} \geqslant \frac{1}{2}\). Since the Shannon entropy \(H(x)\) is monotone decreasing for \(x\geqslant\frac{1}{2}\), we obtain \(H(p_0^{\text{ss}})-H(\epsilon)\geqslant 0\) for \(\epsilon\geqslant\frac{1}{2}\). Combining a similar analysis for the case \(\epsilon< \frac{1}{2}\), we have \(H(p_0^{\text{ss}})-H(\epsilon)\geqslant 0\).}. The first term in Eq.~\eqref{DcpSigma} is the rate of change of the Shannon entropy of the tape, corresponding to the increase in the tape entropy caused by the label interchanges. The second term can be interpreted as a potential thermodynamic cost associated with restoring the tape from the output distribution \(p_0\) to the original input distribution \(\epsilon\). For example, according to the construction of Barato and Seifert~\cite{BaratoSeifert2014PRE}, \(D_{\text{KL}}(p_0^{\text{ss}}\|\epsilon)\) represents the dissipation of an auxiliary system that restores the output distribution to the original distribution.

We point out that the original RNEMD scheme corresponds to an infinite information affinity, \(|F^\psi|\to\infty\), because its interchange operation is deterministic, i.e., \(\epsilon=0\) or \(1\). This results in an infinite Kullback--Leibler divergence and thus an infinite total entropy production rate. For Barato and Seifert's example, this divergence reflects the infinite thermodynamic cost of restoring the tape to its original input distribution.

\section{Steady-state solution and linear response}
In the steady state, Eqs.~\eqref{GvrnEqL}--\eqref{GvrnEqR} for the three regions, subject to the boundary conditions in Eq.~\eqref{GvrnEqB}, reduce to
\begin{align}
    &\partial_x^2\rho_A = \kappa_l^2\left( \rho_A-\rho_l^* \right),\quad x\in[0,a],\label{SSGvrnEqL}\\
    &\partial_x^2\rho_A=0,\quad x\in(a,L-a),\label{SSGvrnEqM}\\
    &\partial_x^2\rho_A = \kappa_r^2\left( \rho_A-\rho_r^* \right),\quad x\in[L-a,L],\label{SSGvrnEqR}
\end{align}
with
\begin{equation}
    \begin{aligned}
        &\kappa_l^2=\frac{\gamma }{D}\frac{\bar{N} + \tau( 2N_A^r - \bar{N})}{2(N_B^lN_A^r+N_A^lN_B^r)},\ \rho_l^*=\frac{\rho(1+\tau)N_A^r}{\bar{N} +\tau(2N_A^r-\bar{N})},\\
        &\kappa_r^2=\frac{\gamma }{D}\frac{\bar{N} - \tau (2N^l_A - \bar{N}) }{2(N_B^lN_A^r+N_A^lN_B^r)},\ \rho_r^*= \frac{\rho(1-\tau)N_A^l}{\bar{N} - \tau(2N_A^l-\bar{N})},\\
        &\tau=\tanh\frac{F^\psi}{2},\qquad \bar{N}=\frac{a}{L}N=N_A^l+N_B^l=N_A^r+N_B^r.
    \end{aligned}
\end{equation}
In this section, we omit the superscript or subscript ``\(\text{ss}\)'' of the steady-state quantities for notational simplicity.

In the following, we assume that \(\kappa_la,\kappa_ra\sim\sqrt{\frac{\gamma a}{D\rho L}}\ll 1\), by using \(N_B^lN_A^r+N_A^lN_B^r\sim(\rho aL)^2\). This condition means that the characteristic diffusion time across a slab, \(t_s^{\text{diff}}=\frac{a^2}{D}\), is much shorter than the characteristic time required to exchange a substantial number of labels in the slabs, \(t_s^{\text{ex}}=\frac{1}{\gamma}\rho aL\). Equivalently, \(\frac{t_s^{\text{diff}}}{t_s^{\text{ex}}} =\frac{\gamma a}{D\rho L}\ll1\). We solve the steady-state equations~\eqref{SSGvrnEqL}--\eqref{SSGvrnEqR} under the boundary conditions in Eq.~\eqref{GvrnEqB} and continuity of the diffusion flux at the slab interfaces \({j_A}_x(a^-)={j_A}_x(a^+)\) and \({j_A}_x(L-a^-)={j_A}_x(L-a^+)\). We also use the definitions \(N_A^l=L\int_0^a\text{d}x\,\rho_A(x)\) and \(N_A^r=L\int_{L-a}^L\text{d}x\,\rho_A(x)\) together with particle-number conservation, \(N_A=L\int_0^L\text{d}x\,\rho_A(x)\). To first order in \(\kappa_la\) and \(\kappa_ra\), the solution is
\begin{equation}\label{rhoAk1}
    \begin{aligned}
        \rho_A(x) =&\ C_l + \rho_l^*,\quad x\in[0,a],\\
        \rho_A(x)=&\ \frac{\rho_r^*-\rho_l^* + C_r-C_l}{L-2a}(x-a) \\
        &+ C_l + \rho_l^*,\quad x\in(a,L-a),\\
        \rho_A(x) =&\ C_r + \rho_r^*,\quad x\in[L-a,L].
    \end{aligned}
\end{equation}
\begin{widetext}
The four coefficients \(N_A^l\), \(N_A^r\), \(C_l\), and \(C_r\) in the solution are determined by
\begin{equation}\label{ClrNlr}
    \begin{aligned}
        &C_l=\frac{N_A^l}{aL}-\rho_l^*,\qquad C_r=\frac{N_A^r}{aL}-\rho_r^*,\\
        &N_A^l=\bar{N}_A+\Delta N_A,\qquad N_A^r=\bar{N}_A-\Delta N_A,
    \end{aligned}
\end{equation}
where \(\bar{N}_A=\frac{a}{L}N_A,\,\bar{N}_B=\frac{a}{L}N_B\), and the particle-number difference \(\Delta N_A\) is determined by the following equation
    \begin{equation}\label{DeltaNA}
        \Delta N_A\left[ \frac{4D}{\gamma a(L-2a)}\left( \Delta N_A^2 + \bar{N}_A\bar{N}_B \right) + \bar{N} \right]=\tanh \frac{F^\psi}{2}\left( \Delta N_A^2 + \bar{N}_A\bar{N}_B \right).
    \end{equation}
\end{widetext}

We focus on the linear-response regime in which the information affinity \(F^\psi\) is weak. Noticing that Eq.~\eqref{DeltaNA} is invariant under \((\Delta N_A,F^\psi)\to(-\Delta N_A,-F^\psi)\), we write the solution as \(\Delta N_A=\frac{1}{2}(\alpha F^\psi + \beta (F^\psi)^3+\dots)\). To first order in \(F^\psi\), the solution of Eq.~\eqref{DeltaNA} is
\begin{equation}
    \Delta N_A=\frac{1}{2}\alpha F^\psi,\qquad\alpha = \frac{\bar{N}_A\bar{N}_B}{\frac{4D}{\gamma a(L-2a)}\bar{N}_A\bar{N}_B + \bar{N}}.
\end{equation}
Substituting this solution into Eq.~\eqref{ClrNlr} and using Eq.~\eqref{rhoAk1}, we obtain the steady-state density distribution \(\rho_A(x)\):
\begin{equation}
    \begin{aligned}
        \rho_A(x) =&\ \frac{N_A}{L^2} + \frac{\alpha}{2aL}F^\psi,\quad x\in[0,a],\\
        \rho_A(x) =&\ \frac{N_A}{L^2} + \frac{\alpha}{2aL}F^\psi \\
        &- \frac{\alpha(x-a)}{aL(L-2a)}F^\psi ,\quad x\in(a,L-a),\\
        \rho_A(x) =&\ \frac{N_A}{L^2} - \frac{\alpha}{2aL}F^\psi,\quad x\in[L-a,L].
    \end{aligned}
\end{equation}
The probability that an output bit is in state 0 can be obtained from this solution and Eq.~\eqref{p0p1} as
\begin{equation}\label{p0}
    p_0 = \frac{1}{2}\left( 1 + \frac{1}{2}F^\psi\frac{\gamma L (L-2a)N}{4DN_AN_B + \gamma L(L-2a)N} \right).
\end{equation}
Substituting this expression into Eq.~\eqref{Jpsi}, we obtain the linear-response relation
\begin{equation}\label{JpsiLin}
    J_A^\psi=L^\psi F^\psi,\qquad L^\psi=\frac{\gamma DN_AN_B}{4DN_AN_B + \gamma L(L-2a)N},
\end{equation}
where \(L^\psi\) is the corresponding linear transport coefficient.

\subsection{Determination of the self-diffusivity}
Generally, the self-diffusivity is determined in RNEMD simulations via Fick's law
\begin{equation}\label{Fick}
    D=-\frac{j_A^{\text{bulk}}}{\partial_x\rho_A^{\text{bulk}}},
\end{equation}
where \(j_A^{\text{bulk}}=\frac{J_A^\psi}{L}\) is the diffusion flux in the bulk region and \(\partial_x\rho_A^{\text{bulk}}\) is the gradient of \(\rho_A\) obtained from its profile in the bulk region. We note that the linear-response relation in Eq.~\eqref{JpsiLin} provides a more direct approach to determining the self-diffusivity:
\begin{equation}\label{DdetLin}
    D=\frac{\gamma L(L-2a)NJ_A^\psi}{N_AN_B(\gamma F^\psi- 4J_A^\psi)},
\end{equation}
where the value of \(J_A^\psi\) can be accurately determined from the net number of A labels exchanged into the left slab, \(N_A^{\text{ex}}\), and the total simulation time, \(t^{\text{tot}}\):
\begin{equation}
    J_A^\psi=\frac{N_A^{\text{ex}}}{t^{\text{tot}}}.
\end{equation}
\begin{figure}[htbp!]
    \centering
    \includegraphics[keepaspectratio, width=0.7\columnwidth]{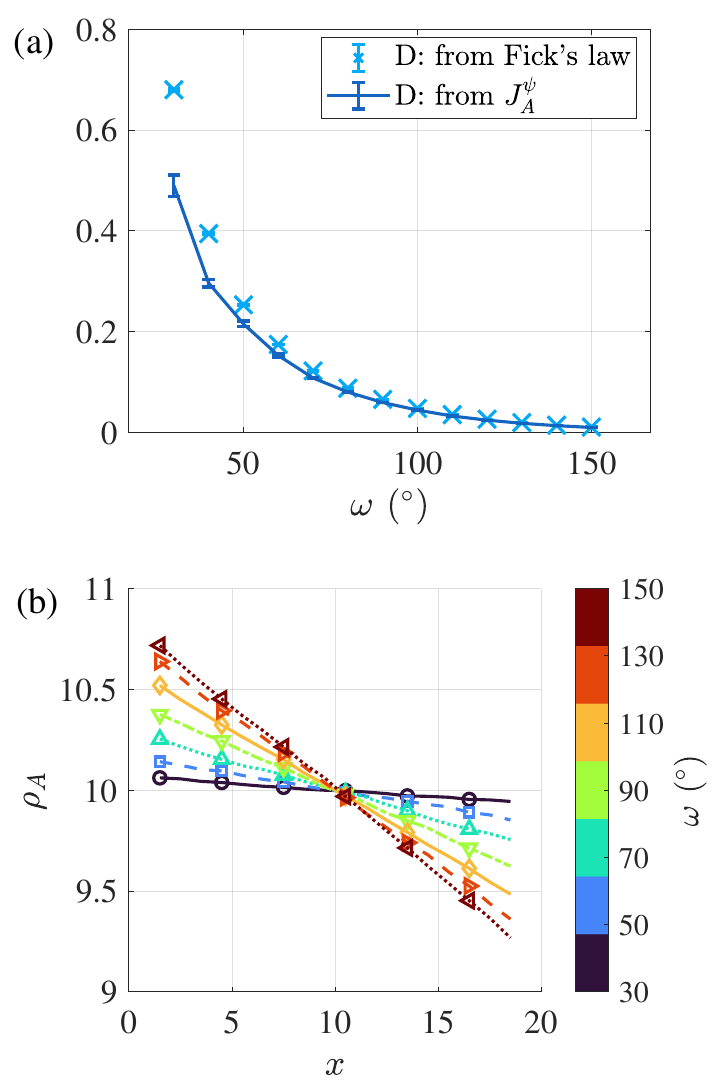}
    \caption{(a) Self-diffusivity measured in RNEMD-SRD simulations at different rotation angles. The solid line represents the values obtained from Fick's law Eq.~\eqref{Fick}, whereas the markers represent those obtained from the linear-response relation in Eq.~\eqref{DdetLin}. (b) Density profiles \(\rho_A\) measured in the simulations. Simulation parameters: \(h=0.1\), \(L=20\), \(a=1\), \(\rho=20\), and \(N_A=N_B=\frac{1}{2}\rho L^2\).}\label{Fig::DRho}
\end{figure}
We verify the linear-response relation~\eqref{JpsiLin} by comparing the values of the self-diffusivity obtained using Fick's law, Eq.~\eqref{Fick}, with those obtained using the linear-response relation, Eq.~\eqref{DdetLin}, in SRD simulations at different rotation angles. As shown in Fig.~\ref{Fig::DRho}(a), the results obtained using the two approaches are in good agreement. At small rotation angles, the self-diffusivity increases significantly, such that the characteristic diffusion length \(\sqrt{Dh}\) of the SRD particles becomes comparable to the slab width \(a=1\). In this regime, the hydrodynamic description is no longer appropriate, leading to the deviations observed in the small-rotation-angle region of Fig.~\ref{Fig::DRho}(a). The corresponding linear density profiles \(\rho_A\) in the bulk region are shown in Fig.~\ref{Fig::DRho}(b).

\section{Macroscopic fluctuations in the steady state}
We have demonstrated that the apparently unphysical reversed force--flux causality in RNEMD can be interpreted within an information-reservoir framework by identifying the information affinity \(F^\psi\) and the resulting current \(J_A^\psi\). In this section, we further investigate steady-state macroscopic fluctuations in RNEMD, including fluctuations of the diffusion fluxes \(\bm{j}_A\) and \(\bm{j}_B\) and the exchange flux \(j_A^\psi\), using macroscopic fluctuation theory. We first construct the large-deviation rate functional for the densities, diffusion fluxes, and exchange flux. Using this functional, we derive a steady-state fluctuation theorem and a thermodynamic uncertainty relation.

\subsection{Large deviation rate functional}
Within the framework of macroscopic fluctuation theory~\cite{Bertini2015}, we describe the system in terms of its densities and currents. We denote the macroscopic state by \(\hat{\Gamma}(t)=(\hat{\rho}_A(\bm{r},t),\hat{\rho}_B(\bm{r},t),\hat{\bm{j}}_A(\bm{r},t),\hat{\bm{j}}_B(\bm{r},t),\hat{j}_A^\psi(\bm{r},\bm{r}',t))\), where a hat denotes a stochastic variable. We focus on the stochastic trajectories of \(\hat{\Gamma}\) on the macroscopic length scale \(L\) and the diffusive time scale \(L^2/D\). The thermodynamic limit is taken as \(N=\rho L^2\to\infty\) with fixed \(\rho\). Specifically, we consider trajectories over the time interval \([0,T_f]\), with the characteristic time \(S_f=\frac{D}{L^2}T_f\) held fixed in this limit.

The evolution of \(\hat{\Gamma}(t)\) satisfies the continuity equations
\begin{equation}\label{conEqs}
    \begin{aligned}
        &\partial_t\hat{\rho}_A = -\bm{\nabla}_{\bm{r}}\cdot\hat{\bm{j}}_A + \Psi[\hat{j}_A^\psi],\\
        &\partial_t\hat{\rho}_B = -\bm{\nabla}_{\bm{r}}\cdot\hat{\bm{j}}_B - \Psi[\hat{j}_A^\psi],
    \end{aligned}
\end{equation}

\begin{equation}
    \Psi[\hat{j}_A^\psi]=\begin{cases}
        \int_{\Omega_r}\text{d}\bm{r}'\hat{j}_A^\psi(\bm{r},\bm{r}',t), & \bm{r}\in\Omega_l,\\
        0, & \bm{r}\in\Omega_m,\\
        -\int_{\Omega_l}\text{d}\bm{r}'\hat{j}_A^\psi(\bm{r}',\bm{r},t), & \bm{r}\in\Omega_r,
    \end{cases}
\end{equation}
and the boundary conditions
\begin{equation}\label{BNDCon}
    \hat{j}_{\xi x}|_{x=0,L}=0,\qquad \xi\in\{A,B\}.
\end{equation}
Averagely, the evolution reduces to Eqs.~\eqref{GvrnEqL}--\eqref{GvrnEqB} with \({\bm{j}}_A=-D\bm{\nabla}_{\bm{r}}{\rho}_A\), \({\bm{j}}_B=-D\bm{\nabla}_{\bm{r}}{\rho}_B\), and \({j}_A^\psi(\bm{r},\bm{r}',t) = \frac{\gamma\epsilon}{Z} \frac{{\rho}_B(\bm{r},t){\rho}_A(\bm{r}',t)}{N_AN_B} - \frac{\gamma(1-\epsilon)}{Z}\frac{{\rho}_A(\bm{r},t){\rho}_B(\bm{r}',t)}{N_AN_B}\) (cf.~Eq.~\eqref{jApsi}).

The probability of observing a given trajectory \(\hat{\Gamma}(t)=\Gamma(t)\) over the time interval \(t\in[0,T_f]\) generally has the large-deviation form~\cite{Bertini2015}
\begin{equation}
    \mathbb{P}\left( \Gamma(t), t\in [0,T_f]\right)\asymp\exp\left[ -N\mathcal{R}_{[0,T_f]}(\Gamma) \right],
\end{equation}
where
\begin{equation}\label{LDR}
    \mathcal{R}_{[0,T_f]}(\Gamma)=\mathcal{V}(\rho_A(\cdot,0),\rho_B(\cdot,0))+ \mathcal{I}_{[0,T_f]}(\Gamma)
\end{equation}
Here,  ``\(\asymp\)'' denotes the logarithmic equivalence as \(N\to\infty\). The functional \(\mathcal{V}(\rho_A(\cdot,0),\rho_B(\cdot,0))\) is the quasipotential of the nonequilibrium steady state and represents the large-deviation rate functional for the initial densities, i.e., \(\mathbb{P}(\rho_A(\cdot,0),\rho_B(\cdot,0))\asymp \exp{(-N\mathcal{V})}\), and \(\mathcal{I}_{[0,T_f]}(\Gamma)\) is the large-deviation rate functional for trajectories starting from \((\rho_A(\cdot,0),\rho_B(\cdot,0))\), i.e., \(\mathbb{P}(\Gamma(t),t\in[0,T_f]\big|(\rho_A(\cdot,0),\rho_B(\cdot,0)))\asymp\exp[-N\mathcal{I}_{[0,T_f]}]\). 

For trajectories satisfying the continuity equations and boundary conditions, the rate functional \(\mathcal{I}_{[0,T_f]}(\Gamma)\) is given by
\begin{widetext}
    \begin{equation}\label{dyLDR}
        \begin{aligned}
            \mathcal{I}_{[0,T_f]}(\Gamma)=& \frac{1}{N} \Bigg\{\int_0^{T_f}\text dt\int_\Omega\text{d}\bm r\left[\frac{\left|\bm j_A+D\bm\nabla_{\bm r}\rho_A\right|^2}{4D\rho_A}+\frac{\left|\bm j_B+D\bm\nabla_{\bm r}\rho_B\right|^2}{4D\rho_B}\right]\\
            &+\int_0^{T_f}\text dt\int_{\Omega_l}\text{d}\bm r\int_{\Omega_r}\text{d}\bm r'\,\mathcal{L}^\psi\left(j_A^\psi(\bm r,\bm r',t);k_+(\bm r,\bm r',t),k_-(\bm r,\bm r',t)\right)\Bigg\},
        \end{aligned}
    \end{equation}
where
\begin{equation}
    \begin{aligned}
        &\mathcal{L}^\psi(j_A^\psi;k_+,k_-)=j_A^\psi\ln\frac{j_A^\psi+\sqrt{{(j_A^\psi)}^2+4k_+k_-}}{2k_+}-\sqrt{{(j_A^\psi)}^2+4k_+k_-}+k_++k_-,
    \end{aligned}
\end{equation}
\begin{equation}\label{kpm}
    k_+(\bm r,\bm r',t)=\frac{\gamma\epsilon}{Z}\frac{\rho_B(\bm r,t)\rho_A(\bm r',t)}{N_AN_B},\qquad k_-(\bm r,\bm r',t)=\frac{\gamma(1-\epsilon)}{Z}\frac{\rho_A(\bm r,t)\rho_B(\bm r',t)}{N_AN_B}.
\end{equation}
For trajectories that violate the continuity equations or boundary conditions, we set \(\mathcal{I}_{[0,T_f]}(\Gamma)=+\infty\).
\end{widetext}
The first term in Eq.~\eqref{dyLDR} describes Gaussian fluctuations of the diffusion fluxes around their expected values specified by the constitutive relations. Such a Gaussian form is valid for many diffusive systems~\cite{Bertini2015}. The second term in Eq.~\eqref{dyLDR} describes Poissonian fluctuations of the exchange flux, and its derivation is provided in Appendix~\ref{APPENDIX::A}. This rate functional has a form similar to that for reaction--diffusion systems~\cite{Bodineau2010}.

\subsection{Large deviation of steady-state entropy production rate}
We further investigate the fluctuations of the steady-state entropy production rate using the rate functional in Eq.~\eqref{LDR}. We define the time-averaged entropy production rate as
\begin{equation}
    \begin{aligned}
        \dot{\hat{\Sigma}}^{\text{tot}}&=\frac{1}{T_f}\int_0^{T_f}\text{d}t\int_{{\Omega}_l}\text{d}\bm{r}\int_{{\Omega}_r}\text{d}\bm{r}'\hat{j}_A^\psi(\bm{r},\bm{r}',t)F^\psi\\
        &\triangleq \hat{J}_A^\psi F^\psi
    \end{aligned}
\end{equation}
whose expectation value in the steady state is \(\dot{\Sigma}_{\text{ss}}^{\text{tot}}\), as given by Eq.~\eqref{SSEPR}. The probability for observing a particular value \(\dot{\hat{\Sigma}}^{\text{tot}}=\dot{{\sigma}}\) of the time-averaged entropy production rate is
\begin{equation}
    \mathbb{P}_{\dot{\hat{\Sigma}}^{\text{tot}}}(\dot{{\sigma}})\asymp\exp[-NS_fI_{\dot{\hat{\Sigma}}^{\text{tot}}}(\dot{{\sigma}})].
\end{equation}
We consider the long-time limit on the diffusive time scale, \(S_f=\frac{D}{L^2}T_f\to\infty\). The contraction principle~\cite{Touchette2009} gives the steady-state large-deviation rate function \(I_{\dot{\hat{\Sigma}}^{\text{tot}}}\) as
\begin{equation}\label{CtrPrnpl}
    I_{\dot{\hat{\Sigma}}^{\text{tot}}}(\dot{{\sigma}})=\lim_{S_f\to\infty}\frac{1}{S_f}\inf_{\Gamma\in\mathcal{A}_{T_f}(\dot{{\sigma}})} \mathcal{R}_{[0,T_f]}(\Gamma)
\end{equation}
where ``\(\inf\)'' represents the infimum, and
\[\begin{aligned}
    &\mathcal{A}_{T_f}(\dot{{\sigma}})= \biggl\{\Gamma(t),t\in[0,T_f]\mid \\
    &\frac{1}{T_f}\int_0^{T_f}\text{d}t\int_{{\Omega}_l}\text{d}\bm{r}\int_{{\Omega}_r}\text{d}\bm{r}'{j}_A^\psi(\bm{r},\bm{r}',t)F^\psi=\dot{{\sigma}}\biggr\}
\end{aligned}\] 
is a set of trajectories that yield the time-averaged entropy production rate \(\dot{{\sigma}}\). We note that \(\dot{\sigma}=\dot{\Sigma}_{\text{ss}}^{\text{tot}}\) is the point at which the rate function attains its minimum value of zero: \(I_{\dot{\hat{\Sigma}}^{\text{tot}}}(\dot{{\sigma}})\geqslant I_{\dot{\hat{\Sigma}}^{\text{tot}}}(\dot{\Sigma}_{\text{ss}}^{\text{tot}})=0\).

The time-reversal properties of the functional \(\mathcal{R}_{[0,T_f]}(\Gamma)\) lead to the Gallavotti--Cohen symmetry of \(I_{\dot{\hat{\Sigma}}^{\text{tot}}}\)~\cite{Bertini2015,Lebowitz1999}, which yields the fluctuation theorem of \(\dot{\hat{\Sigma}}^{\text{tot}}\). We define the time-reversed trajectory \(\Gamma^\dagger\) as \(\Gamma^\dagger=(\rho_A^\dagger(\bm{r},t),\rho_B^\dagger(\bm{r},t),\bm{j}_A^\dagger(\bm{r},t),\bm{j}_B^\dagger(\bm{r},t),(j_A^\psi)^\dagger(\bm{r},\bm{r}',t)),t\in[0,T_f]\), where \(\rho_\xi^\dagger(\bm{r},t)=\rho_\xi(\bm{r},T_f-t)\), \(\bm{j}_\xi^\dagger(\bm{r},t) =-\bm{j}_\xi(\bm{r},T_f-t)\), \(\xi=A,B\), and \((j_A^\psi)^\dagger(\bm{r},\bm{r}',t) =-j_A^\psi(\bm{r},\bm{r}',T_f-t)\). The difference between the functionals for the reversed and original trajectories is
\begin{equation}\label{BFdiffOfR0}
    \begin{aligned}
        &\mathcal{R}(\Gamma^\dagger)-\mathcal{R}(\Gamma) \\
        &=\frac{1}{N}\int_0^{T_f}\text{d}t\Bigg[ -\sum_{\xi\in\{A,B\}}\int_{\Omega}\text{d}\bm{r} \bm{j}_{\xi}\cdot\bm{\nabla}_{\bm{r}}\ln\frac{\rho_\xi}{\rho} \\
        &+ \int_{\Omega_l}\text{d}\bm{r}\int_{\Omega_r}\text{d}\bm{r}'j_A^\psi(\bm{r},\bm{r}',t)\ln\frac{k_+(\bm{r},\bm{r}',t)}{k_-(\bm{r},\bm{r}',t)}\Bigg]+\Delta\mathcal{V},
    \end{aligned}
\end{equation}
where \(\Delta\mathcal{V}=\mathcal{V}|_{t=T_f}-\mathcal{V}|_{t=0}\) is the change in the quasipotential. Then we integrate the first term in the square brackets of Eq.~\eqref{BFdiffOfR0} by parts with the help of the conservation equations~\eqref{conEqs} and the boundary condition~\eqref{BNDCon}, and substitute Eq.~\eqref{kpm} into the second term, which yields:
\begin{equation}\label{BFdiffOfR}
    \mathcal{R}(\Gamma^\dagger)-\mathcal{R}(\Gamma) = \frac{1}{N}T_f \dot{{\sigma}} - \Delta\mathcal{G} + \Delta\mathcal{V},
\end{equation}
where \(\Delta\mathcal{G}\) is the corresponding change of the functional
\begin{equation}
    \begin{aligned}
        \mathcal{G}(\rho_A(\cdot,t),\rho_B(\cdot,t))=\frac{1}{N}\int_{\Omega}\text{d}\bm{r} \sum_{\xi\in\{A,B\}}\rho_\xi\ln\frac{\rho_\xi}{\rho}.
    \end{aligned}
\end{equation}
Equation~\eqref{BFdiffOfR} gives the Gallavotti--Cohen symmetry of \(I_{\dot{\hat{\Sigma}}^{\text{tot}}}\) as follows:
\begin{equation}\label{GCdrv}
    \begin{aligned}
        &I_{\dot{\hat{\Sigma}}^{\text{tot}}}(-\dot{\sigma}) \\
        &= \lim_{S_f\to\infty} \frac1{S_f} \inf_{\Gamma\in\mathcal{A}_{T_f}(-\dot{\sigma})} \mathcal{R}_{[0,T_f]}(\Gamma)\\
        &= \lim_{S_f\to\infty} \frac{1}{S_f} \inf_{\Gamma\in\mathcal{A}_{T_f}(\dot{\sigma})} \mathcal{R}_{[0,T_f]}(\Gamma^\dagger)\\
        &= \frac{1}{\rho D}\dot{\sigma} + \lim_{S_f\to\infty} \frac{1}{S_f} \inf_{\Gamma\in\mathcal{A}_{T_f}(\dot{\sigma})} \mathcal{R}_{[0,T_f]}(\Gamma)\\
        &= \frac{1}{\rho D}\dot{\sigma} + I_{\dot{\hat{\Sigma}}^{\text{tot}}}(\dot{\sigma}).
    \end{aligned}
\end{equation}
Here the second equality comes from the fact that \(\dot{{\sigma}}\) is odd under time reversal, and the sets \(\mathcal{A}_{T_f}(\dot{{\sigma}})\) and \(\mathcal{A}_{T_f}(-\dot{{\sigma}})\) can be mapped onto each other under time reversal. The third equality follows from Eq.~\eqref{BFdiffOfR} and the limits \(\lim_{S_f\to\infty}\frac{\Delta\mathcal{G}}{S_f}=0,\,\lim_{S_f\to\infty}\frac{\Delta\mathcal{V}}{S_f}=0\), since the boundary terms \(\Delta\mathcal{G}\) and \(\Delta\mathcal{V}\) are of order \(O(1)\) in the long-time limit. Equivalently, Eq.~\eqref{GCdrv} reads,
\begin{equation}\label{GCsym}
    I_{\dot{\hat{\Sigma}}^{\text{tot}}}(-\dot{\sigma}) - I_{\dot{\hat{\Sigma}}^{\text{tot}}}(\dot{\sigma}) = \frac{1}{\rho D}\dot{\sigma}.
\end{equation}
This symmetry yields the steady-state fluctuation theorem for the entropy production rate:
\begin{equation}\label{FT}
    \frac{\mathbb{P}_{\dot{\hat{\Sigma}}^{\text{tot}}}(\dot{{\sigma}})}{\mathbb{P}_{\dot{\hat{\Sigma}}^{\text{tot}}}(-\dot{{\sigma}})}\asymp e^{T_f\dot{{\sigma}}}.
\end{equation}

The fluctuation theorem is a general relation that has been proved in a wide class of systems including those subjected to deterministic and stochastic dynamics~\cite{Evans1993,Gallavotti1995,Jorge1998,Lebowitz1999} and is valid arbitrarily far from equilibrium. Typically, near equilibrium, it yields the fluctuation-dissipation theorem and the Onsager reciprocal relations~\cite{Gallavotti1996,Jorge1998,Lebowitz1999}. Consequently, the fluctuation theorem Eq.~\eqref{FT} derived here supports the thermodynamic consistency of the information-reservoir interpretation of RNEMD at the level of macroscopic fluctuations.

\subsection{Steady-state thermodynamic uncertainty relation}
From the contraction principle in Eq.~\eqref{CtrPrnpl}, we derive a bound on the large deviation rate function for the steady-state entropy-production, from which the thermodynamic uncertainty relation~\cite{BaratoSeifert2015,Gingrich2016} follows. Our derivation follows the large-deviation approach of Gingrich et al.~\cite{Gingrich2016}. Consider a typical steady-state trajectory \(\Gamma_{\text{ss}}(t)=(\rho_A^{\text{ss}},\rho_B^{\text{ss}},\bm{j}_A^{\text{ss}},\bm{j}_B^{\text{ss}},j_{A,\text{ss}}^\psi),\,t\in[0,T_f]\). Herein, \(\rho_A^{\text{ss}}\) satisfies the steady-state equations~\eqref{SSGvrnEqL}-\eqref{SSGvrnEqR}, \(\rho_A^{\text{ss}}=\rho-\rho_B^{\text{ss}}\), and \(\bm{j}_A^{\text{ss}},\bm{j}_B^{\text{ss}},j_{A,\text{ss}}^\psi\) are given by
\begin{equation}
    \begin{aligned}
        &\bm{j}_\xi^{\text{ss}}=-D\bm{\nabla}_{\bm{r}}\rho_\xi^{\text{ss}},\qquad \xi\in\{A,B\}\\
        &j_{A,\text{ss}}^\psi =k_+^{\text{ss}}-k_-^{\text{ss}},
    \end{aligned}
\end{equation}
where \(k_+^{\text{ss}}\) and \(k_-^{\text{ss}}\) are obtained by replacing \(\rho_A\) and \(\rho_B\) in Eq.~\eqref{kpm} with \(\rho_A^{\text{ss}}\) and \(\rho_B^{\text{ss}}\), respectively. The time-averaged entropy production rate along this typical trajectory is
\begin{equation}
    \begin{aligned}
        \dot{\Sigma}_{\text{ss}}^{\text{tot}}&=J^\psi_{A,\text{ss}}F^\psi,\\
        J^\psi_{A,\text{ss}}&=\int_{{\Omega}_l}\text{d}\bm{r}\int_{{\Omega}_r}\text{d}\bm{r}'j_{A,\text{ss}}^\psi(\bm{r},\bm{r}').
    \end{aligned}
\end{equation}

We now introduce an auxiliary trajectory \(\Gamma_\theta\) by scaling the fluxes along the trajectory \(\Gamma_{\text{ss}}\) by an arbitrary constant \(\theta\in\mathbb{R}\):
\begin{equation}
    \Gamma_\theta=(\rho_A^{\text{ss}},\rho_B^{\text{ss}},\theta\bm{j}_A^{\text{ss}},\theta\bm{j}_B^{\text{ss}},\theta j_{A,\text{ss}}^\psi).
\end{equation}
This auxiliary trajectory still satisfies the steady-state evolutions [Eq.~\eqref{conEqs} with \(\partial_t\hat{\rho}_A=\partial_t\hat{\rho}_B=0\)] but deviates from the constitutive relation. The auxiliary trajectory yields the following entropy production rate
\begin{equation}
    \dot{\Sigma}_\theta=\theta\dot{\Sigma}_{\text{ss}}^{\text{tot}}.
\end{equation}
By substituting \(\Gamma=\Gamma_\theta\) into the functional \(\mathcal{R}_{[0,T_f]}[\Gamma]\) in Eq.~\eqref{CtrPrnpl}, using Eq.~\eqref{dyLDR}, and exploiting the time independence and \(y\)-translational invariance of the densities and currents of the auxiliary trajectory, we transform the equality in Eq.~\eqref{CtrPrnpl} into the following inequality:
\begin{widetext}
\begin{equation}
    \begin{aligned}
        I_{\dot{\hat{\Sigma}}^{\text{tot}}}(\dot{\Sigma}_\theta)&\leqslant \frac{(\theta-1)^2}{4\rho D}\int_{\Omega}\text{d}\bm{r}\left( -{{j}_A^{\text{ss}}}_x\partial_x\ln\frac{\rho_A^{\text{ss}}}{\rho_B^{\text{ss}}} \right) + \frac{1}{\rho D}\int_{\Omega_l}\text{d}\bm r\int_{\Omega_r}\text{d}\bm r'\,\mathcal{L}^\psi\left(\theta j_{A,\text{ss}}^\psi(\bm r,\bm r',t);k_+^{\text{ss}}(\bm r,\bm r',t),k_-^{\text{ss}}(\bm r,\bm r',t)\right)
    \end{aligned}
\end{equation}
Using the following inequality, whose proof is provided in Appendix~\ref{APPENDIX::B},
\begin{equation}\label{IneqTbPf}
    \mathcal{L}^\psi\left(\theta j_{A,\text{ss}}^\psi;k_+^{\text{ss}},k_-^{\text{ss}}\right)\leqslant \frac{(\theta-1)^2}{4} j_{A,\text{ss}}^\psi\ln\frac{k_+^{\text{ss}}}{k_-^{\text{ss}}},
\end{equation}
and Eqs.~\eqref{EPRdiff}, \eqref{EPRpsiori}-\eqref{SSEPR0}, we obtain
\begin{equation}
    \begin{aligned}
        I_{\dot{\hat{\Sigma}}^{\text{tot}}}(\dot{\Sigma}_\theta)&\leqslant \frac{(\theta-1)^2}{4\rho D}\left\{ \int_{\Omega}\text{d}\bm{r}\left( -{{j}_A^{\text{ss}}}_x\partial_x\ln\frac{\rho_A^{\text{ss}}}{\rho_B^{\text{ss}}} \right) + \int_{\Omega_l}\text{d}\bm r\int_{\Omega_r}\text{d}\bm r'\, j_{A,\text{ss}}^\psi(\bm r,\bm r',t)\ln\frac{k_+^{\text{ss}}(\bm r,\bm r',t)}{k_-^{\text{ss}}(\bm r,\bm r',t)} \right\}\\
        &=\frac{(\theta-1)^2}{4\rho D}\left( \dot{\Sigma}^{\text{diff}}_{\text{ss}} + \dot{\Sigma}^{\psi}_{\text{ss}} \right)=\frac{(\theta-1)^2}{4\rho D}\dot{\Sigma}_{\text{ss}}^{\text{tot}},
    \end{aligned}
\end{equation}
\end{widetext}
i.e.,
\begin{equation}\label{genBnd}
    I_{\dot{\hat{\Sigma}}^{\text{tot}}}(\dot{\Sigma}_\theta)\leqslant\frac{(\dot{\Sigma}_\theta-\dot{\Sigma}_{\text{ss}}^{\text{tot}})^2}{4\rho D\dot{\Sigma}_{\text{ss}}^{\text{tot}}}.
\end{equation}
Since both sides of inequality~\eqref{genBnd} attain their common minimum value of zero at
\(\dot{\Sigma}_\theta=\dot{\Sigma}_{\text{ss}}^{\text{tot}}\), we obtain
\begin{equation}\label{genBnd2}
    I_{\dot{\hat{\Sigma}}^{\text{tot}}}''(\dot{\Sigma}_{\text{ss}}^{\text{tot}})\leqslant\frac{1}{2\rho D\dot{\Sigma}_{\text{ss}}^{\text{tot}}}.
\end{equation}
We then use the relation \(I_{\dot{\hat{\Sigma}}^{\text{tot}}}''(\dot{\Sigma}_{\text{ss}}^{\text{tot}})=\frac{1}{NS_f\text{Var}_{ss}(\dot{\hat{\Sigma}}^{\text{tot}})}\)~\footnote{This relation follows by expanding the rate function around its minimum, \(I_{\dot{\hat{\Sigma}}^{\text{tot}}}(\dot{\Sigma}_\theta)\simeq \frac{1}{2}(\dot{\Sigma}_\theta-\dot{\Sigma}_{\text{ss}}^{\text{tot}})^2I_{\dot{\hat{\Sigma}}^{\text{tot}}}''(\dot{\Sigma}_{\text{ss}}^{\text{tot}})\), and comparing the resulting probability distribution \(\mathbb{P}_{\dot{\hat{\Sigma}}^{\text{tot}}}\asymp \exp{[-NS_fI_{\dot{\hat{\Sigma}}^{\text{tot}}}]}\) with a Gaussian distribution of variance \(\text{Var}_{\text{ss}}(\dot{\hat{\Sigma}}^{\text{tot}})\).},
where \(\text{Var}_{\text{ss}}(\dot{\hat{\Sigma}}^{\text{tot}})\) is the variance of the steady-state entropy production rate.  Using \(NS_f=\rho DT_f\), inequality~\eqref{genBnd2} becomes
\begin{equation}\label{genBnd3}
    \frac{1}{T_f\text{Var}_{ss}(\dot{\hat{\Sigma}}^{\text{tot}})}\leqslant\frac{1}{2\dot{\Sigma}_{\text{ss}}^{\text{tot}}}.
\end{equation}
Substituting \(\dot{\Sigma}_{\text{ss}}^{\text{tot}}=J_{A,\text{ss}}^\psi F^\psi\) and \(\text{Var}_{ss}(\dot{\hat{\Sigma}}^{\text{tot}})=(F^\psi)^2\text{Var}_{\text{ss}}(\hat{J}_A^\psi)\) into inequality~\eqref{genBnd3}, we arrive at the thermodynamic uncertainty relation in the thermodynamic and long-time limits:
\begin{equation}\label{TUR}
    \frac{\text{Var}_{\text{ss}}(\hat{J}_A^{\psi})}{(J_{A,\text{ss}}^\psi)^2}\Sigma_{\text{ss}}^{\text{tot}}\geqslant 2,
\end{equation}
where \(\Sigma_{\text{ss}}^{\text{tot}}=T_f\dot{\Sigma}_{\text{ss}}^{\text{tot}}\) is the total steady-state entropy production. This uncertainty relation bounds the precision of the exchange current by the total entropy production. Since \(\Sigma_{\text{ss}}^{\text{tot}}=T_f|J_{A,\text{ss}}^\psi||F^\psi|\), it connects transport precision to the information affinity. Moreover, the thermodynamic uncertainty relation can also be understood in terms of fluctuation-response relation in the general nonequilibrium steady state~\cite{Dechant2020}. In the present case, Eq.~\eqref{TUR} can be rearranged as 
\begin{equation}
    |J_{A,\text{ss}}^\psi|\leqslant\frac{1}{2}T_f\text{Var}_{\text{ss}}(\hat{J}_A^{\psi})|F^\psi|,
\end{equation}
which provides an inequality relating the response of the exchange current to the information affinity and the steady-state fluctuations of the exchange current, and is reminiscent of the fluctuation-dissipation relation near equilibrium.

\begin{figure}[htbp!]
    \centering
    \includegraphics[keepaspectratio, width=\columnwidth]{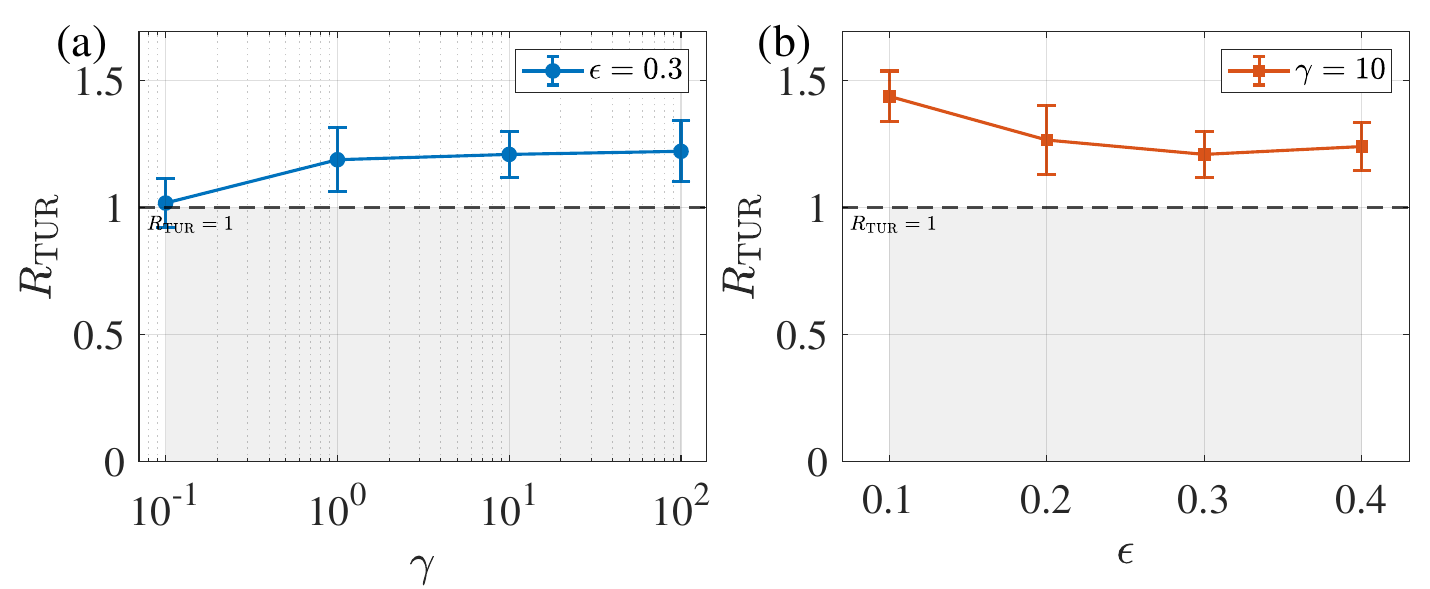}
    \caption{Thermodynamic uncertainty relation ratio \(R_{\text{TUR}}\) computed in the simulations for different values of \(\gamma\) and \(\epsilon\). (a) Results for \(\gamma=0.1,\,1,\,10,\,100\) at fixed \(\epsilon=0.3\). (b) Results for \(\epsilon=0.1,\,0.2,\,0.3,\,0.4\) at fixed \(\gamma=10\). The other parameters are \(h=0.1\), \(L=20\), \(a=1\), \(\rho=20\), \(N_A=N_B=\frac{1}{2}\rho L^2\), and \(\omega=2\pi/3\).}\label{Fig::TUR}
\end{figure}

We verify the thermodynamic uncertainty relation~\eqref{TUR} using RNEMD-SRD simulations. We calculate the thermodynamic uncertainty ratio defined by \(R_{\text{TUR}}\triangleq\frac{\text{Var}_{\text{ss}}(\hat{J}_A^{\psi})\Sigma_{\text{ss}}^{\text{tot}}}{2(J_{A,\text{ss}}^\psi)^2}\) for different values of the parameters \(\gamma\) and \(\epsilon\). As shown in Fig.~(\ref{Fig::TUR}), the simulation results are consistent with the thermodynamic uncertainty relation \(R_{\text{TUR}}\geqslant 1\) within the error bars.

\section{Conclusion}
We have interpreted the apparently unphysical reversal of force--flux causality in label-exchange RNEMD within an information-reservoir framework. The bias of the incoming bit sequence defines an information affinity \(F^\psi\) that drives the exchange current \(J_A^\psi\). Within the hydrodynamic description, we found that the macroscopic entropy production rate takes the flux--affinity form \(\dot\Sigma^{\mathrm{tot}}=J_A^\psi F^\psi\). In the linear-response regime, we obtained the steady-state solution and the transport coefficient relating \(F^\psi\) to \(J_A^\psi\). This relation provides an alternative way to determine the self-diffusivity, and our simulations show agreement with estimates based on Fick's law within the regime where the hydrodynamic description applies. Moreover, we investigate the steady-state macroscopic fluctuations through the large deviation rate functional of the density-current trajectory. With the help of this rate functional, we further derive a fluctuation theorem of the total entropy production and a thermodynamic uncertainty relation.

The information-reservoir viewpoint may also provide a starting point for analyzing other RNEMD schemes, including those used to generate shear flows or temperature gradients. Establishing such extensions would require identifying the relevant information variables and exchange rules for each scheme. Beyond RNEMD, other reverse nonequilibrium simulation methods are also expected to be interpreted within the framework of information thermodynamics. More broadly, this viewpoint may help guide the design of other versatile NEMD methods.

\section*{Acknowledgment}
We acknowledge the support from the National Natural Science Foundation of China (Grant No. T2325027).

\appendix
\section{Derivation of the density–current large-deviation functional}\label{APPENDIX::A}
We first introduce expressions for the empirical densities and currents. The empirical density \(\hat{\rho}_\xi(\bm{r},t)\) and particle current \(\hat{\bm{j}}_\xi(\bm{r},t)\) of particles labeled \(\xi\in\{A,B\}\) are defined by
\begin{equation}\label{EmRho}
    \hat{\rho}_\xi(\bm{r},t)=\sum_i\delta(\hat{\bm{r}}_i(t)-\bm{r})\delta_{\xi_i(t),\xi},
\end{equation}
\begin{equation}\label{EmJD}
    \hat{\bm{j}}_\xi(\bm{r},t)=\sum_i\hat{\dot{\bm{r}}}_i(t)\delta(\hat{\bm{r}}_i(t)-\bm{r})\delta_{\xi_i(t),\xi},
\end{equation}
where \(\hat{\bm{r}}_i(t)\), \(\hat{\dot{\bm{r}}}_i(t)\), and \(\xi_i(t)\) are the position, velocity, and label of the particle \(i\) at time \(t\), respectively, and \(\delta_{\xi_i(t),\xi}\) is the Kronecker delta. 

To express the empirical exchange current \(\hat{j}_A^\psi\), we first define \(\mathcal{E}_+\) as the set of \((B,A)\to(A,B)\) exchange events during the observation interval \([0,T_f]\), and denote the number of events in \(\mathcal{E}_+\) by \(\hat{\mathcal{N}}_+\). We use \(\hat{t}_e\in[0,T_f]\) to denote the time of event \(e\in\mathcal{E}_+\), and \(\hat{\bm{r}}_{e,l}\in\Omega_l\) and \(\hat{\bm{r}}_{e,r}\in\Omega_r\) to denote the positions of the particles in the left and the right slabs, respectively. The empirical exchange current of A particles flowing from \(\bm{r}'\in\Omega_r\) to \(\bm{r}\in\Omega_l\) is therefore expressed as
\begin{equation}\label{EmJPsiP}
    \hat{k}_+(\bm{r},\bm{r}',t)=\sum_{e\in\mathcal{E}_+}\delta(\bm{r}-\hat{\bm{r}}_{e,l})\delta(\bm{r}'-\hat{\bm{r}}_{e,r})\delta(t-\hat{t}_e).
\end{equation}
The empirical exchange current of \(A\) particles flowing from \(\bm{r}\in\Omega_l\) to \(\bm{r}'\in\Omega_r\) is similarly defined as
\begin{equation}\label{EmJPsiM}
    \hat{k}_-(\bm{r},\bm{r}',t)=\sum_{e\in\mathcal{E}_-}\delta(\bm{r}-\hat{\bm{r}}_{e,l})\delta(\bm{r}'-\hat{\bm{r}}_{e,r})\delta(t-\hat{t}_e),
\end{equation}
where \(\mathcal{E}_-\) is the set of \((A,B)\to(B,A)\) exchange events during \(t\in[0,T_f]\), and \(\hat{\mathcal{N}}_-\) is the number of such events. Combining Eqs.~\eqref{EmJPsiP} and \eqref{EmJPsiM}, we define the net empirical exchange current as
\begin{equation}
    \hat{j}_A^\psi(\bm{r},\bm{r}',t)=\hat{k}_+(\bm{r},\bm{r}',t) - \hat{k}_-(\bm{r},\bm{r}',t),\quad \bm{r}\in\Omega_l,\,\bm{r}'\in\Omega_r.
\end{equation}

One can confirm that the empirical densities and currents satisfy the continuity equations
\begin{equation}\label{EmConEq}
    \begin{aligned}
        &\partial_t\hat{\rho}_A = -\bm{\nabla}_{\bm{r}}\cdot\hat{\bm{j}}_A + \Psi[\hat{j}_A^\psi],\\
        &\partial_t\hat{\rho}_B = -\bm{\nabla}_{\bm{r}}\cdot\hat{\bm{j}}_B - \Psi[\hat{j}_A^\psi],
    \end{aligned}
\end{equation}
where
\begin{equation}
    \Psi[\hat{j}_A^\psi]=\begin{cases}
        \int_{\Omega_r}\text{d}\bm{r}'\hat{j}_A^\psi(\bm{r},\bm{r}',t), & \bm{r}\in\Omega_l,\\
        0, & \bm{r}\in\Omega_m,\\
        -\int_{\Omega_l}\text{d}\bm{r}'\hat{j}_A^\psi(\bm{r}',\bm{r},t), & \bm{r}\in\Omega_r,
    \end{cases}
\end{equation}
together with the boundary conditions
\begin{equation}
    {\hat{j}_{Ax}}|_{x=0,L}={\hat{j}_{Bx}}|_{x=0,L}=0.
\end{equation}

We focus on density and current fluctuations on the macroscopic length scale \(L\) and the characteristic diffusive time scale \(L^2/D\). It is therefore convenient to introduce the rescaled coordinate and time as
\begin{equation}\label{rescaledVar}
    \bm{R}=\frac{\bm{r}}{L},\qquad s=\frac{D}{L^2}t.
\end{equation}
We denote the rescaled observation interval by \([0,S_f],\,S_f=\frac{D}{L^2}T_f\). Rescaling the continuity equation~\eqref{EmConEq} yields
\begin{equation}
    \begin{aligned}
        &\partial_s\hat{n}_A = -\bm{\nabla}_{\bm{R}}\cdot\hat{\bm{u}}_A + \tilde{\Psi}[\hat{u}_A^\psi],\\
        &\partial_s\hat{n}_B = -\bm{\nabla}_{\bm{R}}\cdot\hat{\bm{u}}_B - \tilde{\Psi}[\hat{u}_A^\psi],
    \end{aligned}
\end{equation}
where
\begin{equation}
    \tilde{\Psi}[\hat{u}_A^\psi]=\begin{cases}
        \int_{\tilde{\Omega}_r}\text{d}\bm{R}'\hat{u}_A^\psi(\bm{R},\bm{R}',s), & \bm{R}\in\tilde{\Omega}_l,\\
        0, & \bm{R}\in\tilde{\Omega}_m,\\
        -\int_{\tilde{\Omega}_l}\text{d}\bm{R}'\hat{u}_A^\psi(\bm{R}',\bm{R},s), & \bm{R}\in\tilde{\Omega}_r,
    \end{cases}
\end{equation}
Here, \(\tilde{\Omega}_l\), \(\tilde{\Omega}_m\), and \(\tilde{\Omega}_r\) are the regions for the rescaled coordinates corresponding to \(\Omega_l\), \(\Omega_m\), and \(\Omega_r\), respectively. The corresponding rescaled densities \(\hat{n}_\xi\), \(\xi\in\{A,B\}\), and currents \(\hat{\bm{u}}_\xi\) and \(\hat{u}_A^\psi\) are
\begin{equation}\label{rescaledDenCur}
    \begin{aligned}
        &\hat{n}_\xi(\bm{R},s)=\hat{\rho}_\xi(L\bm{R},\frac{L^2}{D}s)/\rho,\\
        &\hat{\bm{u}}_\xi(\bm{R},s)=\frac{L}{\rho D}\hat{\bm{j}}_\xi(L\bm{R},\frac{L^2}{D}s),\\
        &\hat{u}_A^\psi(\bm{R},\bm{R}',s)=\frac{L^4}{\rho D}\hat{j}_A^\psi(L\bm{R},L\bm{R}',\frac{L^2}{D}s),
    \end{aligned}
\end{equation}
According to Eqs.~\eqref{EmRho}--\eqref{EmJPsiM}, these rescaled densities and currents are given by
\begin{equation}
    \hat{n}_\xi(\bm{R},s)=\frac{1}{N}\sum_i\delta(\hat{\bm{R}}_i(s)-\bm{R})\delta_{\xi_i(s),\xi},
\end{equation}
\begin{equation}
    \hat{\bm{u}}_\xi(\bm{R},s)=\frac{1}{N}\sum_i\hat{\dot{\bm{R}}}_i(s)\delta(\hat{\bm{R}}_i(s)-\bm{R})\delta_{\xi_i(s),\xi},
\end{equation}
and
\begin{equation}
    \hat{u}_A^\psi(\bm{R},\bm{R}',s)=\hat{\tilde{k}}_+(\bm{R},\bm{R}',s)-\hat{\tilde{k}}_-(\bm{R},\bm{R}',s),
\end{equation}
with
\begin{equation}\label{REMjpsiPM}
    \hat{\tilde{k}}_{\pm}(\bm{R},\bm{R}',s)=\frac{1}{N}\sum_{e\in\mathcal{E}_{\pm}}\delta(\bm{R}-\hat{\bm{R}}_{e,l})\delta(\bm{R}'-\hat{\bm{R}}_{e,r})\delta(s-\hat{s}_e).
\end{equation}
Note that the expected number of exchange events \(\hat{\mathcal{N}}_{\pm}\) during \(s\in[0,S_f]\) (i.e., the number of terms in the summation of Eq.~\eqref{REMjpsiPM}) is of order \(N\): \(\mathbb{E}[\hat{\mathcal{N}}_{\pm}]\sim\gamma T_f=\frac{\gamma S_f}{\rho D}N\).

Consider a trajectory \[\hat{\tilde{\Gamma}}(s)=(\hat{n}_A(s),\hat{n}_B(s),\hat{\bm{u}}_A(s),\hat{\bm{u}}_B(s),\hat{u}_A^\psi(s)),\,s\in[0,S_f]\] with the initial densities \((\hat{n}_A(s),\hat{n}_B(s))|_{s=0}=({n}_{A0},{n}_{B0})\). The large-deviation form of the probability of observing a given trajectory \(\hat{\tilde{\Gamma}}(s)={\tilde{\Gamma}}(s)\) is
\begin{equation}
    \begin{aligned}
        &\mathbb{P}(\tilde{\Gamma}(s),s\in[0,S_f]\mid({n}_{A0},{n}_{B0}))\\
        &\asymp\exp[-N\mathcal{I}_{[0,S_f]}({n}_A,{n}_B,{\bm{u}}_A,{\bm{u}}_B,{u}_A^\psi)].
    \end{aligned}
\end{equation}
If the trajectory \(\tilde{\Gamma}(s)\) violates the continuity equations or the prescribed boundary conditions, the rate functional is infinite:
\begin{equation}
    \mathcal{I}_{[0,S_f]}=+\infty.
\end{equation}

We consider that diffusion and label exchange are driven by independent noise. The rate functional \(\mathcal{I}_{[0,S_f]}\) can therefore be decomposed into a diffusion part and an exchange part:
\begin{equation}\label{TotLDR}
    \begin{aligned}
        &\mathcal{I}_{[0,S_f]}({n}_A,{n}_B,{\bm{u}}_A,{\bm{u}}_B,{u}_A^\psi)\\
        &=\mathcal{I}^{\text{diff}}_{[0,S_f]}(n_A,n_B,\bm{u}_A,\bm{u}_B)+\mathcal{I}^\psi_{[0,S_f]}(n_A,n_B, u_A^\psi).
    \end{aligned}
\end{equation}
The diffusion part \(\mathcal{I}^{\text{diff}}_{[0,S_f]}\) is given by
\begin{equation}\label{DiffLDR}
    \begin{aligned}
        &\mathcal{I}^{\text{diff}}_{[0,S_f]}(n_A,n_B,\bm{u}_A,\bm{u}_B)\\
        =&\ \frac{1}{4}\int_0^{S_f}\text{d}s\int_{\tilde{\Omega}}\text{d}\bm{R}\left( \frac{1}{n_A}\left\lvert \bm{u}_A+\bm{\nabla}_{\bm{R}}n_A \right\rvert^2 \right.\\
        &\left.+ \frac{1}{n_B}\left\lvert \bm{u}_B+\bm{\nabla}_{\bm{R}}n_B \right\rvert^2 \right),
    \end{aligned}
\end{equation}
which describes the Gaussian fluctuations of the diffusion currents.

To derive the exchange part of the rate functional in Eq.~\eqref{TotLDR}, we discretize space into \(N_d\times N_d\) small volume elements and the time interval \([0,S_f]\) into \(N_d\) small intervals. We denote the volume element \(\bm{R}\) by \(\Delta\bm{R}\), and denote the starting point and length of the \(k\)-th (\(k=1,\dots,N_d\)) time interval by \(s_k\) and \(\Delta s_k\), respectively. We define \(\hat{N}_{+,\bm{R},\bm{R}',s_k}\) as the number of the label-exchange events between a \(B\) particle in the volume element \(\bm{R}\) of the left slab and an \(A\) particle in the volume element \(\bm{R}'\) of the right slab during the time interval \(s_k\):
\begin{equation}
    \begin{aligned}
        \hat{N}_{+,\bm{R},\bm{R}',s_k}&=\sum_{e\in\mathcal{E}_+}\delta_{\bm{R},[\hat{\bm{R}}_{e,l}]}\delta_{\bm{R}',[\hat{\bm{R}}_{e,r}]}\delta_{s_k,[\hat{s}_e]}\\
        &\triangleq N\hat{u}_{+,\bm{R},\bm{R}',s_k}^\psi\Delta\bm{R}\Delta\bm{R}'\Delta s_k.
    \end{aligned}
\end{equation}
Here, \([\hat{\bm{R}}_{e,l}]\), \([\hat{\bm{R}}_{e,r}]\), and \([\hat{s}_e]\) represent the labels of the volume elements and time intervals corresponding to \(\hat{\bm{R}}_{e,l}\), \(\hat{\bm{R}}_{e,r}\), and \(\hat{s}_e\), respectively. We have introduced the discretized exchange current \(\hat{u}_{+,\bm{R},\bm{R}',s_k}^\psi\), whose continuum limit is \(\hat{u}_+^\psi(\bm{R},\bm{R}',s)\), i.e.,
\begin{equation}
    \begin{aligned}
        &\lim_{N_d\to\infty}\hat{u}_{+,\bm{R},\bm{R}',s_k}^\psi\\
        &=\lim_{N_d\to\infty}\frac{\sum_{e\in\mathcal{E}_+}\delta_{\bm{R},[\hat{\bm{R}}_{e,l}]}\delta_{\bm{R}',[\hat{\bm{R}}_{e,r}]}\delta_{s_k,[\hat{s}_e]}}{N\Delta\bm{R}\Delta\bm{R}'\Delta s_k}\\
        &=\frac{1}{N}\sum_{e\in\mathcal{E}_+}\delta(\hat{\bm{R}}_{e,l}-\bm{R})\delta(\hat{\bm{R}}_{e,r}-\bm{R}')\delta(s-\hat{s}_e)\\
        &=\hat{u}_+^\psi(\bm{R},\bm{R}',s).
    \end{aligned}
\end{equation}

Over each short coarse-grained time interval, we take the densities as their values at the beginning of the interval. Under this approximation, the numbers of exchange events can be modeled as Poisson random variables. Thus, the number of events \(\hat{N}_{+,\bm{R},\bm{R}',s_k}\) follows a Poisson distribution with the conditional expectation
\begin{equation}\label{aveNp}
    \begin{aligned}
        &\mathbb{E}[\hat{N}_{+,\bm{R},\bm{R}',s_k}\mid (n_{A,s_k}, n_{B,s_k})]\\
        &=\frac{\gamma\epsilon L^2}{Z D}\tilde{p}(\bm{R},B;\bm{R}',A,s_k)\Delta\Lambda\\
        &=N\frac{\gamma\epsilon }{\rho D Z}\tilde{p}(\bm{R},B;\bm{R}',A,s_k)\Delta\Lambda\\
        &=N\frac{\gamma\epsilon }{\rho D Z}\frac{N^2}{N_AN_B}n_B(\bm{R},s_k)n_A(\bm{R}',s_k)\Delta\Lambda\\
        &\triangleq N\tilde{k}_{+,\bm{R},\bm{R}',s_k}\Delta\Lambda,
    \end{aligned}
\end{equation}
where \(\Delta\Lambda=\Delta\bm{R}\Delta\bm{R}'\Delta s_k\), \(Z=\frac{N_B^lN_A^r+N_A^lN_B^r}{N_AN_B}\) (cf. Eq.~\eqref{jApsi}), \(\tilde{p}(\bm{R},B;\bm{R}',A,s_k)\) is the rescaled two-particle distribution function, and \((n_{A,s_k}, n_{B,s_k})\) denotes the densities at the beginning of the time interval.
In the third equality, we have expressed the two-particle distribution in terms of the densities by using the single-particle distributions (cf. Eq.~\eqref{pBA}). The last equality defines the rescaled mean local exchange current of A into the left slab, \(\tilde{k}_{+,\bm{R},\bm{R}',s_k}\) (cf. Eqs.~\eqref{jApsiDef}--\eqref{pBA}). Similarly, we define the number of exchange events in the opposite direction as \(\hat{N}_{-,\bm{R},\bm{R}',s_k}=\sum_{e\in\mathcal{E}_-}\delta_{\bm{R},[\hat{\bm{R}}_{e,l}]}\delta_{\bm{R}',[\hat{\bm{R}}_{e,r}]}\delta_{s_k,[\hat{s}_e]}\triangleq N\hat{u}_{-,\bm{R},\bm{R}',s_k}^\psi\Delta\Lambda,\) whose conditional expectation reads
\begin{equation}\label{aveNm}
    \begin{aligned}
        &\mathbb{E}[\hat{N}_{-,\bm{R},\bm{R}',s_k} \mid (n_{A,s_k}, n_{B,s_k})]\\
        &=N\frac{\gamma(1-\epsilon) }{\rho D Z}\tilde{p}(\bm{R},A;\bm{R}',B,s_k)\Delta\Lambda\\
        &=N\frac{\gamma(1-\epsilon) }{\rho D Z}\frac{N^2}{N_AN_B}n_A(\bm{R},s_k)n_B(\bm{R}',s_k)\Delta\Lambda\\
        &\triangleq N\tilde{k}_{-,\bm{R},\bm{R}',s_k}\Delta\Lambda.
    \end{aligned}
\end{equation}
Defining the net number of exchange events as
\begin{equation}
    \hat{N}_{\text{net},\bm{R},\bm{R}',s_k}=\hat{N}_{+,\bm{R},\bm{R}',s_k}-\hat{N}_{-,\bm{R},\bm{R}',s_k},
\end{equation}
we obtain the total discretized exchange current:
\begin{equation}\label{uApsi_Net}
    \begin{aligned}
        \hat{u}_{A,\bm{R},\bm{R}',s_k}^\psi&=\hat{u}_{+,\bm{R},\bm{R}',s_k}^\psi-\hat{u}_{-,\bm{R},\bm{R}',s_k}^\psi\\
        &=\frac{\hat{N}_{\text{net},\bm{R},\bm{R}',s_k}}{N\Delta\Lambda}.
    \end{aligned}
\end{equation}

We write the the probability \(P({u}_{A,\bm{R},\bm{R}',s_k}^\psi\mid (n_{A,s_k},n_{B,s_k}))\) in large deviation form as
\begin{equation}
    P({u}_{A,\bm{R},\bm{R}',s_k}^\psi\mid (n_{A,s_k},n_{B,s_k}))\asymp \exp{\left[ -NI_{\bm{R},\bm{R}',s_k} \right]}.
\end{equation}
According to the Gärtner--Ellis theorem~\cite{Touchette2009}, the rate function \(I_{\bm{R},\bm{R}',s_k}\) can be obtained through the Legendre--Fenchel transform of the scaled cumulant generating function (SCGF) of \(u_{A,\bm{R},\bm{R}',s_k}^\psi\):
\begin{equation}\label{LFTran}
    I_{\bm{R},\bm{R}',s_k}=\sup_{z}\{zu_{A,\bm{R},\bm{R}',s_k}^\psi-G(z)\},
\end{equation}
where ``\(\sup\)'' represents the supremum. Here \(G(z)\) is the SCGF of \({u}_{A,\bm{R},\bm{R}',s_k}^\psi\), defined by
\begin{equation}
    G(z)=\lim_{N\to\infty}\frac{1}{N}\ln \mathbb{E}[\exp{Nz\hat{u}_{A,\bm{R},\bm{R}',s_k}^\psi}\mid (n_{A,s_k},n_{B,s_k})].
\end{equation}
Substituting Eq.~\eqref{uApsi_Net} in this definition and introducing the cumulant generating function (CGF) of \(\hat{N}_{\text{net},\bm{R},\bm{R}',s_k}\) as
\begin{equation}
    \begin{aligned}
        G_{N_{\text{net}}}(z)&=\ln\mathbb{E}[\exp{[ z\hat{N}_{\text{net},\bm{R},\bm{R}',s_k} ]}\mid (n_{A,s_k},n_{B,s_k})],
    \end{aligned}
\end{equation}
we obtain the SCGF of \(u_{A,\bm{R},\bm{R}',s_k}^\psi\) as
\begin{equation}\label{Gz}
    G(z)=\lim_{N\to\infty}\frac{1}{N}G_{N_{\text{net}}}(\frac{z}{\Delta\Lambda}).
\end{equation}
Because \(\hat{N}_{+,\bm{R},\bm{R}',s_k}\) and \(\hat{N}_{-,\bm{R},\bm{R}',s_k}\) are Poisson random variables with the expectation values given by Eqs~\eqref{aveNp} and \eqref{aveNm}, respectively, the CGF \(G_{N_{\text{net}}}(z)\) is
\begin{equation}
    G_{N_{\text{net}}}(z) = N[(e^z-1)\tilde{k}_{+} + (e^{-z}-1)\tilde{k}_{-}]\Delta\Lambda.
\end{equation}
Equation~\eqref{Gz} subsequently yields the following SCGF:
\begin{equation}
    \begin{aligned}
        G(z)=&\ \left[ (e^{\frac{z}{\Delta\Lambda}}-1)\tilde{k}_{+,\bm{R},\bm{R}',s_k} \right.\\
        &\left.+ (e^{-\frac{z}{\Delta\Lambda}}-1)\tilde{k}_{-,\bm{R},\bm{R}',s_k} \right]\Delta\Lambda.
    \end{aligned}
\end{equation}
Finally, substituting this expression for \(G(z)\) into Eq.~\eqref{LFTran}, we obtain the rate function
\begin{widetext}
\begin{equation}\label{LresInLFT}
    \begin{aligned}
        I_{\bm{R},\bm{R}',s_k} &=\sup_{z}\left\{zu_{A,\bm{R},\bm{R}',s_k}^\psi-\left[ (e^{z}-1)\tilde{k}_{+,\bm{R},\bm{R}',s_k} + (e^{-z}-1)\tilde{k}_{-,\bm{R},\bm{R}',s_k} \right]\right\}\Delta\Lambda\\
        &= \mathcal{L}^\psi(u_{A,\bm{R},\bm{R}',s_k}^\psi,\tilde{k}_{+,\bm{R},\bm{R}',s_k},\tilde{k}_{-,\bm{R},\bm{R}',s_k})\Delta\Lambda,
    \end{aligned}
\end{equation}
where the function \(\mathcal{L}^\psi(u_A^\psi,\tilde{k}_+,\tilde{k}_-)\) is
\begin{equation}
    \begin{aligned}
        \mathcal{L}^\psi(u_A^\psi,\tilde{k}_+,\tilde{k}_-)=&\ u_A^\psi\ln\frac{u_A^\psi+\sqrt{(u_A^\psi)^2+4\tilde{k}_+\tilde{k}_-}}{2\tilde{k}_+} - \sqrt{(u_A^\psi)^2+4\tilde{k}_+\tilde{k}_-} + \tilde{k}_+ + \tilde{k}_-.
    \end{aligned}
\end{equation}

The exchange contribution to the joint density--current rate functional is obtained by summing \(I_{\bm{R},\bm{R}',s_k}\) along the trajectory. In the continuum limit we obtain
 \begin{equation}\label{ExcLDR}
    \begin{aligned}
        \mathcal{I}^\psi_{[0,S_f]}(n_A,n_B, u_A^\psi)&=\lim_{N_d\to\infty}\sum_{\bm{R},\bm{R}',k}I_{\bm{R},\bm{R}',s_k}=\lim_{N_d\to\infty}\sum_{\bm{R},\bm{R}',k}\mathcal L^\psi(u_A^\psi;\tilde k_+,\tilde k_-)\Delta\Lambda\\
        &=\int_0^{S_f}\text{d}s\int_{\tilde{\Omega}_l}\text{d}\bm{R}\int_{\tilde{\Omega}_r}\text{d}\bm{R}' \mathcal{L}^\psi(u_{A}^\psi(\bm{R},\bm{R}',s),\tilde{k}_{+}(\bm{R},\bm{R}',s),\tilde{k}_{-}(\bm{R},\bm{R}',s)).
    \end{aligned}
\end{equation}

Returning to the original dimensional variables in both contributions to the rate functional, Eqs.~\eqref{DiffLDR} and \eqref{ExcLDR}, using Eqs.~\eqref{rescaledVar} and \eqref{rescaledDenCur}, we obtain the rate functional given by Eq.~\eqref{dyLDR} in the main text:
\begin{equation}
\begin{aligned}
        \mathcal{I}_{[0,T_f]}(\Gamma)=& \frac{1}{N} \Bigg\{\int_0^{T_f}\text dt\int_\Omega\text{d}\bm r\left[\frac{\left|\bm j_A+D\bm\nabla_{\bm r}\rho_A\right|^2}{4D\rho_A}+\frac{\left|\bm j_B+D\bm\nabla_{\bm r}\rho_B\right|^2}{4D\rho_B}\right]\\
        &+\int_0^{T_f}\text dt\int_{\Omega_l}\text{d}\bm r\int_{\Omega_r}\text{d}\bm r'\,\mathcal{L}^\psi\left(j_A^\psi(\bm r,\bm r',t);k_+(\bm r,\bm r',t),k_-(\bm r,\bm r',t)\right)\Bigg\},
    \end{aligned}
\end{equation}

\end{widetext}

\section{Proof of the inequality~\texorpdfstring{\eqref{IneqTbPf}}{}}\label{APPENDIX::B}
In this appendix, we prove the following inequality
\begin{equation}\label{SimIneqTbPf}
    \mathcal{L}^\psi(u)\leqslant\frac{1}{4j}(u-j)^2F,
\end{equation}
where
\begin{equation}
    \begin{aligned}
        &\mathcal{L}^\psi(u) \\
        &= u\ln\frac{u+\sqrt{u^2+4ab}}{2a} - \sqrt{u^2+4ab} + a + b,
    \end{aligned}
\end{equation}
and
\begin{equation}
    j=a-b,\qquad F=\ln\frac{a}{b},
\end{equation}
with \(a,b>0\). Setting \(a=k_+^{\text{ss}},b=k_-^{\text{ss}}, j=j_{A,\text{ss}}^\psi,u=\theta j_{A,\text{ss}}^\psi\), inequality~\eqref{SimIneqTbPf} becomes inequality~\eqref{IneqTbPf} used in the main text.

The Legendre--Fenchel transform of \(\mathcal{L}^\psi(u)\) is
\begin{equation}
    \begin{aligned}
        \overline{\mathcal{L}^\psi}(z)&=\sup_u\left\{ zu-\mathcal{L}^\psi(u) \right\}\\
        &=ae^z+be^{-z}-(a+b)\\
        &=\sqrt{ab}\left( \sqrt{\frac{a}{b}}e^z + \sqrt{\frac{b}{a}}e^{-z} - \sqrt{\frac{a}{b}} - \sqrt{\frac{b}{a}} \right)\\
        &=2\sqrt{ab}\left[ \cosh(z+\frac{1}{2}F) -\cosh\frac{F}{2} \right],
    \end{aligned}
\end{equation}
where we have used \(F=\ln\frac{a}{b}\) in the last equality. We denote the right-hand side of the inequality~\eqref{SimIneqTbPf} by \[\mathcal{L}_u(u)=\frac{1}{4j}(u-j)^2F,\] whose Legendre-Fenchel transform is
\begin{equation}
    \begin{aligned}
        \overline{\mathcal{L}_u}(z)&=\sup_u\left\{ zu-\mathcal{L}_u(u) \right\}\\
        &=\frac{zj}{F}(z+F)\\
        &=2\sqrt{ab}\frac{z}{F}(z+F)\sinh(\frac{1}{2}F),
    \end{aligned}
\end{equation}
where we have used \(j=a-b\) and \(F=\ln\frac{a}{b}\) in the last equality. Proving inequality~\eqref{SimIneqTbPf} between \(\mathcal{L}^\psi(u)\) and \(\mathcal{L}_u(u)\) can be reduced to proving the following inequality between their Legendre--Fenchel transforms:
\begin{equation}\label{ineqOfduals}
    \overline{\mathcal{L}^\psi}(z)\geqslant \overline{\mathcal{L}_u}(z).
\end{equation}
This is because the original inequality~\eqref{SimIneqTbPf} can be obtained by using the inequality~\eqref{ineqOfduals} and the inverse transforms \(\mathcal{L}^\psi(u)=\sup_z\{uz-\overline{\mathcal{L}^\psi}(z)\}\leqslant\sup_z\{uz-\overline{\mathcal{L}_u}(z)\}=\mathcal{L}_u(u)\). 

We define the difference between the two sides of inequality~\eqref{ineqOfduals} as
\begin{equation}
    g(z)=\overline{\mathcal{L}^\psi}(z)-\overline{\mathcal{L}_u}(z),
\end{equation}
which can be written as
\begin{widetext}
\begin{equation}
    g(z)= 2\sqrt{ab}\left[ \cosh(z+\frac{1}{2}F) -\cosh\frac{F}{2}- \frac{z}{F}(z+F)\sinh\frac{F}{2} \right].
\end{equation}
We note \(z=0\) and \(z=-F\) are zeros of \(g(z)\). The derivative of \(g(z)\) is
\begin{equation}
    \begin{aligned}
        g'(z) &= 2\sqrt{ab}\left( z+\frac{1}{2}F \right)\left[ f\left( z+\frac{1}{2}F \right) - f\left( \frac{1}{2}F \right) \right],
    \end{aligned}
\end{equation}
\end{widetext}
where \(f(x)=\frac{\sinh x}{x}\) is monotonically increasing for \(x>0\) and monotonically decreasing for \(x<0\). The function \(g'(z)\) has zeros at \[z=0,\qquad z=-\frac{F}{2},\qquad z=-F.\] Using the monotonicity of \(f(x)\), we find that the sign of \(g'(z)\) changes from negative to positive as \(z\) increases through \(z=0\) and \(z=-F\).  It follows that \(z=0\) and \(z=-F\) are global minima of \(g(z)\). Therefore,
\begin{equation}
    g(z)\geqslant 0,
\end{equation}
which proves inequality~\eqref{ineqOfduals}.

\bibliographystyle{apsrev}
\bibliography{RefList}

@article{BaratoSeifert2014PRE,
  author  = {Barato, Andre C. and Seifert, Udo},
  title   = {Stochastic Thermodynamics with Information Reservoirs},
  journal = {Physical Review E},
  volume  = {90},
  number  = {4},
  pages   = {042150},
  year    = {2014},
  month   = oct,
  doi     = {10.1103/PhysRevE.90.042150},
  url     = {https://doi.org/10.1103/PhysRevE.90.042150}
}

@article{BaratoSeifert2014PRL,
  author  = {Barato, Andre C. and Seifert, Udo},
  title   = {Unifying Three Perspectives on Information Processing in Stochastic Thermodynamics},
  journal = {Physical Review Letters},
  volume  = {112},
  number  = {9},
  pages   = {090601},
  year    = {2014},
  month   = mar,
  doi     = {10.1103/PhysRevLett.112.090601},
  url     = {https://doi.org/10.1103/PhysRevLett.112.090601}
}

@article{DeffnerJarzynski2013,
  author  = {Deffner, Sebastian and Jarzynski, Christopher},
  title   = {Information Processing and the Second Law of Thermodynamics: An Inclusive, Hamiltonian Approach},
  journal = {Physical Review X},
  volume  = {3},
  number  = {4},
  pages   = {041003},
  year    = {2013},
  month   = oct,
  doi     = {10.1103/PhysRevX.3.041003},
  url     = {https://doi.org/10.1103/PhysRevX.3.041003}
}

@article{HorowitzEsposito2014,
  author  = {Horowitz, Jordan M. and Esposito, Massimiliano},
  title   = {Thermodynamics with Continuous Information Flow},
  journal = {Physical Review X},
  volume  = {4},
  number  = {3},
  pages   = {031015},
  year    = {2014},
  month   = jul,
  doi     = {10.1103/PhysRevX.4.031015},
  url     = {https://doi.org/10.1103/PhysRevX.4.031015}
}

@article{MandalJarzynski2012,
  author  = {Mandal, Dibyendu and Jarzynski, Christopher},
  title   = {Work and Information Processing in a Solvable Model of {Maxwell's} Demon},
  journal = {Proceedings of the National Academy of Sciences},
  volume  = {109},
  number  = {29},
  pages   = {11641--11645},
  year    = {2012},
  month   = jul,
  doi     = {10.1073/pnas.1204263109},
  url     = {https://doi.org/10.1073/pnas.1204263109}
}

@incollection{MullerPlatheBordat2004,
  author    = {M{\"u}ller-Plathe, Florian and Bordat, Patrice},
  title     = {Reverse Non-equilibrium Molecular Dynamics},
  booktitle = {Novel Methods in Soft Matter Simulations},
  editor    = {Karttunen, Mikko and Lukkarinen, Ari and Vattulainen, Ilpo},
  series    = {Lecture Notes in Physics},
  volume    = {640},
  pages     = {310--326},
  publisher = {Springer},
  address   = {Berlin, Heidelberg},
  year      = {2004},
  doi       = {10.1007/978-3-540-39895-0_10},
  url       = {https://doi.org/10.1007/978-3-540-39895-0_10}
}

@article{MullerPlathe1999,
  author  = {M{\"u}ller-Plathe, Florian},
  title   = {Reversing the Perturbation in Nonequilibrium Molecular Dynamics: An Easy Way to Calculate the Shear Viscosity of Fluids},
  journal = {Physical Review E},
  volume  = {59},
  number  = {5},
  pages   = {4894--4898},
  year    = {1999},
  month   = may,
  doi     = {10.1103/PhysRevE.59.4894},
  url     = {https://doi.org/10.1103/PhysRevE.59.4894}
}

@article{Muller1997,
    author = {Müller-Plathe, Florian},
    title = {A simple nonequilibrium molecular dynamics method for calculating the thermal conductivity},
    journal = {The Journal of Chemical Physics},
    volume = {106},
    number = {14},
    pages = {6082-6085},
    year = {1997},
    month = {04},
    issn = {0021-9606},
    doi = {10.1063/1.473271},
    url = {https://doi.org/10.1063/1.473271}
}

@article{SagawaUeda2010,
  author  = {Sagawa, Takahiro and Ueda, Masahito},
  title   = {Generalized {Jarzynski} Equality under Nonequilibrium Feedback Control},
  journal = {Physical Review Letters},
  volume  = {104},
  number  = {9},
  pages   = {090602},
  year    = {2010},
  month   = mar,
  doi     = {10.1103/PhysRevLett.104.090602},
  url     = {https://doi.org/10.1103/PhysRevLett.104.090602}
}

@article{SagawaUeda2012,
  author  = {Sagawa, Takahiro and Ueda, Masahito},
  title   = {Fluctuation Theorem with Information Exchange: Role of Correlations in Stochastic Thermodynamics},
  journal = {Physical Review Letters},
  volume  = {109},
  number  = {18},
  pages   = {180602},
  year    = {2012},
  month   = nov,
  doi     = {10.1103/PhysRevLett.109.180602},
  url     = {https://doi.org/10.1103/PhysRevLett.109.180602}
}

@article{ParrondoHorowitzSagawa2015,
  author  = {Parrondo, Juan M. R. and Horowitz, Jordan M. and Sagawa, Takahiro},
  title   = {Thermodynamics of Information},
  journal = {Nature Physics},
  volume  = {11},
  number  = {2},
  pages   = {131--139},
  year    = {2015},
  month   = feb,
  doi     = {10.1038/nphys3230},
  url     = {https://doi.org/10.1038/nphys3230}
}

@book{NET,
  author    = {S. R. De Groot, P. Mazur},
  title     = {Non-Equilibrium Thermodynamics},
  publisher = {Dover Publications},
  year      = {2011},
  month     = jan,
  isbn      = {9780486647418}
}

@book{EvansMorriss2008,
  author    = {Evans, Denis J. and Morriss, Gary P.},
  title     = {Statistical Mechanics of Nonequilibrium Liquids},
  edition   = {2},
  publisher = {Cambridge University Press},
  address   = {Cambridge},
  year      = {2008},
  doi       = {10.1017/CBO9780511535307},
  isbn      = {978-0-521-85791-8}
}

@article{HeffelfingerSwol1994,
    author = {Heffelfinger, Grant S. and Swol, Frank van},
    title = {Diffusion in Lennard-Jones fluids using dual control volume grand canonical molecular dynamics simulation (DCV-GCMD)},
    journal = {The Journal of Chemical Physics},
    volume = {100},
    number = {10},
    pages = {7548-7552},
    year = {1994},
    month = {05},
    issn = {0021-9606},
    doi = {10.1063/1.466849},
    url = {https://doi.org/10.1063/1.466849}
}

@book{Maxwell, 
  place={Cambridge}, 
  series={Cambridge Library Collection - Physical Sciences}, 
  title={Theory of Heat}, 
  publisher={Cambridge University Press}, 
  author={Maxwell, James Clerk}, 
  year={2011}, 
  collection={Cambridge Library Collection - Physical Sciences}
}

@article{MPC1,
    author = {Malevanets, Anatoly and Kapral, Raymond},
    title = "{Mesoscopic model for solvent dynamics}",
    journal = {The Journal of Chemical Physics},
    volume = {110},
    number = {17},
    pages = {8605-8613},
    year = {1999},
    month = {05}
}

@article{Gompper_2009,
  author    = "Gompper, G. and Ihle, T. and Kroll, D. M. and Winkler, R. G.",
  title     = "Multi-Particle Collision Dynamics: A Particle-Based Mesoscale Simulation Approach to the Hydrodynamics of Complex Fluids",
  journal   = "Adv. Polym. Sci.",
  year      = "2009",
  volume    = {221},
  pages     = {1},
}

@article{MPC_RS,
  title = {Stochastic rotation dynamics: A Galilean-invariant mesoscopic model for fluid flow},
  author = {Ihle, T. and Kroll, D. M.},
  journal = {Phys. Rev. E},
  volume = {63},
  issue = {2},
  pages = {020201},
  numpages = {4},
  year = {2001},
  month = {Jan},
  publisher = {American Physical Society}
}

@article{SRD_GK1,
  title = {Stochastic rotation dynamics. I. Formalism, Galilean invariance, and Green-Kubo relations},
  author = {Ihle, T. and Kroll, D. M.},
  journal = {Phys. Rev. E},
  volume = {67},
  issue = {6},
  pages = {066705},
  numpages = {11},
  year = {2003},
  month = {Jun},
  publisher = {American Physical Society}
}

@article{MPC_MD,
  title = {Hydrodynamic interactions and Brownian forces in colloidal suspensions: Coarse-graining over time and length scales},
  author = {Padding, J. T. and Louis, A. A.},
  journal = {Phys. Rev. E},
  volume = {74},
  issue = {3},
  pages = {031402},
  numpages = {29},
  year = {2006},
  month = {Sep},
  publisher = {American Physical Society}
}

@article{BelushkinLiviFoffi2011,
  title = {Hydrodynamics and the Fluctuation Theorem},
  author = {Belushkin, M. and Livi, R. and Foffi, G.},
  journal = {Phys. Rev. Lett.},
  volume = {106},
  issue = {21},
  pages = {210601},
  numpages = {4},
  year = {2011},
  month = {May},
  publisher = {American Physical Society},
  doi = {10.1103/PhysRevLett.106.210601},
  url = {https://link.aps.org/doi/10.1103/PhysRevLett.106.210601}
}

@article{Luo2020,
  title = {Onsager reciprocal relations with broken time-reversal symmetry},
  author = {Luo, Rongxiang and Benenti, Giuliano and Casati, Giulio and Wang, Jiao},
  journal = {Phys. Rev. Res.},
  volume = {2},
  issue = {2},
  pages = {022009(R)},
  numpages = {4},
  year = {2020},
  month = {Apr},
  publisher = {American Physical Society},
  doi = {10.1103/PhysRevResearch.2.022009},
  url = {https://link.aps.org/doi/10.1103/PhysRevResearch.2.022009}
}

@article{Schnakenberg1976,
  title = {Network theory of microscopic and macroscopic behavior of master equation systems},
  author = {Schnakenberg, J.},
  journal = {Rev. Mod. Phys.},
  volume = {48},
  issue = {4},
  pages = {571--585},
  numpages = {0},
  year = {1976},
  month = {Oct},
  publisher = {American Physical Society},
  doi = {10.1103/RevModPhys.48.571},
  url = {https://link.aps.org/doi/10.1103/RevModPhys.48.571}
}

@article{Bertini2015,
  title = {Macroscopic fluctuation theory},
  author = {Bertini, Lorenzo and De Sole, Alberto and Gabrielli, Davide and Jona-Lasinio, Giovanni and Landim, Claudio},
  journal = {Rev. Mod. Phys.},
  volume = {87},
  issue = {2},
  pages = {593--636},
  numpages = {44},
  year = {2015},
  month = {Jun},
  publisher = {American Physical Society},
  doi = {10.1103/RevModPhys.87.593},
  url = {https://link.aps.org/doi/10.1103/RevModPhys.87.593}
}

@article{Bodineau2010,
  author  = {Bodineau, T. and Lagouge, M.},
  title   = {Current Large Deviations in a Driven Dissipative Model},
  journal = {Journal of Statistical Physics},
  year    = {2010},
  volume  = {139},
  number  = {2},
  pages   = {201--218},
  doi     = {10.1007/s10955-010-9934-7},
  issn    = {1572-9613},
  url     = {https://doi.org/10.1007/s10955-010-9934-7}
}

@article{Touchette2009,
title = {The large deviation approach to statistical mechanics},
journal = {Physics Reports},
volume = {478},
number = {1},
pages = {1-69},
year = {2009},
issn = {0370-1573},
doi = {https://doi.org/10.1016/j.physrep.2009.05.002},
url = {https://www.sciencedirect.com/science/article/pii/S0370157309001410},
author = {Hugo Touchette}
}

@article{Lebowitz1999,
  author  = {Lebowitz, Joel L. and Spohn, Herbert},
  title   = {A Gallavotti–Cohen-Type Symmetry in the Large Deviation Functional for Stochastic Dynamics},
  journal = {Journal of Statistical Physics},
  year    = {1999},
  volume  = {95},
  number  = {1},
  pages   = {333--365},
  month   = apr,
  doi     = {10.1023/A:1004589714161},
  url     = {https://doi.org/10.1023/A:1004589714161},
  issn    = {1572-9613}
}

@article{Evans1993,
  title = {Probability of second law violations in shearing steady states},
  author = {Evans, Denis J. and Cohen, E. G. D. and Morriss, G. P.},
  journal = {Phys. Rev. Lett.},
  volume = {71},
  issue = {15},
  pages = {2401--2404},
  numpages = {0},
  year = {1993},
  month = {Oct},
  publisher = {American Physical Society},
  doi = {10.1103/PhysRevLett.71.2401},
  url = {https://link.aps.org/doi/10.1103/PhysRevLett.71.2401}
}

@article{Gallavotti1995,
  author  = {Gallavotti, G. and Cohen, E. G. D.},
  title   = {Dynamical ensembles in stationary states},
  journal = {Journal of Statistical Physics},
  year    = {1995},
  volume  = {80},
  number  = {5},
  pages   = {931--970},
  month   = sep,
  doi     = {10.1007/BF02179860},
  url     = {https://doi.org/10.1007/BF02179860},
  issn    = {1572-9613}
}

@article{Jorge1998,
doi = {10.1088/0305-4470/31/16/003},
url = {https://doi.org/10.1088/0305-4470/31/16/003},
year = {1998},
month = {apr},
publisher = {},
volume = {31},
number = {16},
pages = {3719},
author = {Jorge Kurchan},
title = {Fluctuation theorem for stochastic dynamics},
journal = {Journal of Physics A: Mathematical and General}
}

@article{Gallavotti1996,
  title = {Extension of Onsager's Reciprocity to Large Fields and the Chaotic Hypothesis},
  author = {Gallavotti, Giovanni},
  journal = {Phys. Rev. Lett.},
  volume = {77},
  issue = {21},
  pages = {4334--4337},
  numpages = {0},
  year = {1996},
  month = {Nov},
  publisher = {American Physical Society},
  doi = {10.1103/PhysRevLett.77.4334},
  url = {https://link.aps.org/doi/10.1103/PhysRevLett.77.4334}
}

@article{Dechant2020,
author = {Andreas Dechant and {S.-i. Sasa} },
title = {Fluctuation–response inequality out of equilibrium},
journal = {Proceedings of the National Academy of Sciences},
volume = {117},
number = {12},
pages = {6430-6436},
year = {2020},
doi = {10.1073/pnas.1918386117},
URL = {https://www.pnas.org/doi/abs/10.1073/pnas.1918386117}
}

@article{Gingrich2016,
  title = {Dissipation Bounds All Steady-State Current Fluctuations},
  author = {Gingrich, Todd R. and Horowitz, Jordan M. and Perunov, Nikolay and England, Jeremy L.},
  journal = {Phys. Rev. Lett.},
  volume = {116},
  issue = {12},
  pages = {120601},
  numpages = {5},
  year = {2016},
  month = {Mar},
  publisher = {American Physical Society},
  doi = {10.1103/PhysRevLett.116.120601},
  url = {https://link.aps.org/doi/10.1103/PhysRevLett.116.120601}
}

@article{BaratoSeifert2015,
  title = {Thermodynamic Uncertainty Relation for Biomolecular Processes},
  author = {Barato, Andre C. and Seifert, Udo},
  journal = {Phys. Rev. Lett.},
  volume = {114},
  issue = {15},
  pages = {158101},
  numpages = {5},
  year = {2015},
  month = {Apr},
  publisher = {American Physical Society},
  doi = {10.1103/PhysRevLett.114.158101},
  url = {https://link.aps.org/doi/10.1103/PhysRevLett.114.158101}
}

\end{document}